\documentclass[english]{IEEEtran}
\usepackage[T1]{fontenc}
\usepackage[latin9]{inputenc}
\usepackage{float}
\usepackage{amsthm}
\usepackage{amsmath}
\usepackage{graphicx}
\usepackage{amssymb}

\makeatletter

\floatstyle{ruled}
\newfloat{algorithm}{tbp}{loa}
\floatname{algorithm}{Algorithm}

\theoremstyle{plain}
\newtheorem{thm}{Theorem}
 \theoremstyle{definition}
  \newtheorem{example}[thm]{Example}
  \theoremstyle{plain}
  \newtheorem{lem}[thm]{Lemma}
  \theoremstyle{definition}
  \newtheorem{defn}[thm]{Definition}
  \theoremstyle{plain}
  \newtheorem{cor}[thm]{Corollary}
  \theoremstyle{remark}
  
  \theoremstyle{remark}
  \newtheorem*{acknowledgement*}{Acknowledgement}

\usepackage{pstricks}
\usepackage{pst-node}
\usepackage{algorithm}
\usepackage[noend]{algpseudocode}
\usepackage{cite}

\@ifundefined{showcaptionsetup}{}{%
 \PassOptionsToPackage{caption=false}{subfig}}
\usepackage{subfig}
\makeatother

\usepackage{babel}

\begin{document}
\global\long\def\tp{\rightsquigarrow_{\pi}}

\global\long\def\cost{\operatorname{cost}}

\global\long\def\queede{\hfill\qed}

\global\long\def\upto{:}

\global\long\def\cutedge#1#2{#1,\!#2\text{{-cut\!\ edge}}}

\title{Sorting of Permutations by Cost-Constrained Transpositions}

\author{Farzad Farnoud (Hassanzadeh) and Olgica Milenkovic, \emph{IEEE Member }}
\maketitle
\begin{abstract}
We address the problem of finding the minimum decomposition of a permutation
in terms of transpositions with non-uniform cost. For arbitrary non-negative
cost functions, we describe polynomial-time, constant-approximation
decomposition algorithms. For metric-path costs, we describe exact
polynomial-time decomposition algorithms. Our algorithms represent
a combination of Viterbi-type algorithms and graph-search techniques
for minimizing the cost of individual transpositions, and dynamic
programing algorithms for finding minimum cost cycle decompositions.
The presented algorithms have applications in information theory,
bioinformatics, and algebra.
\end{abstract}

\section{Introduction}

Permutations are ubiquitous combinatorial objects encountered in areas
as diverse as mathematics, computer science, communication theory,
and bioinformatics. The set of all permutations of $n$ elements --
the symmetric group of order $n$, $\mathbb{S}_{n}$ -- plays an important
role in algebra, representation theory, and analysis of algorithms
\cite{goulden_combinatorial_2004,van_lint_course_2001,chung_algebraic_????,hofri_analysis_1995}.
As a consequence, the properties of permutations and the symmetric
group have been studied extensively.

One of the simplest ways to generate an arbitrary permutation is to
apply a sequence of transpositions - swaps of two elements - on a
given permutation, usually the identity permutation. The sequence
of swaps can be reversed in order to recover the identity permutation
from the original permutation. This process is  referred to as sorting
by transpositions. 

A simple result, established by Cayley in the 1860's, asserts that
the minimum number of transpositions needed to sort a permutation
so as to obtain the identity permutation is the difference of the
size of the permutation and the number of cycles formed by the elements
of the permutation. Cayley's result is based on a simple constructive
argument, which reduces to a linear-complexity procedure for breaking
cycles into sub-cycles. Sorting a permutation is equivalent to finding
the transposition distance between the permutation and the identity
permutation. Since permutations form a group, the transposition distance
between two arbitrary permutations equals the transposition distance
between the identity permutation and the composition of the inverse
of one permutation and the other permutation.

We address the substantially more challenging question: assuming that
each transposition has a non-negative, but otherwise arbitrary cost,
is it possible to find the minimum sorting cost and the sequence of
transpositions used for this sorting in polynomial time? In other
words, can one compute the cost-constrained transposition distance
between two permutations in polynomial time? Although at this point
it is not known if the problem is NP hard, at first glance, it appears
to be computationally difficult, due to the fact that it is related
to finding minimum generators of groups and the subset-sum problem\cite{jerrum_complexity_1985}.
Nevertheless, we show that large families of cost functions -- such
as costs based on metric-paths -- have exact polynomial-time decomposition
algorithms. Furthermore, we devise algorithms for approximating the
minimum sorting cost for any non-negative cost function, with an approximation
constant that does not exceed four.

Our investigation is motivated by three different applications. 

The first application pertains to sorting of genomic sequences, while
the second application is related to a generalization of the notion
of a chemical channel (also known as trapdoor channel \cite{permuter_capacity_2008}).
The third application is in the area of coding for storage devices.

Genomic sequences -- such as DNA sequences -- evolved from one common
ancestor, and therefore frequently contain conserved subsequences.
During evolution or during the onset of a genomic disease, these subsequences
are subject to mutations, and they may exchange their locations. As
an example, genomes of cancer cells tend to contain the same sequence
of blocks as normal cells, but in a reshuffled (permuted) order. This
finding motivated a large body of work on developing efficient algorithms
for reverse-engineering the sequence of shuffling steps performed
on conserved subsequences. With a few exceptions, most of the methods
for sorting use reversals rather than transpositions, they follow
the uniform cost model (each change in the ordering of the blocks
is equally likely) and the most parsimonious sorting scenario (the
sorting scenario with smallest number of changes is the most likely
explanation for the observed order). Several approaches that do not
fit into this framework were described in~\cite{pinter_genomic_2002,pevzner_transforming_2003}.
Sorting by cost-constrained transpositions can be seen as a special
instant of the general subsequence sorting problem,where the sequence
is allowed to break at three or four points. Unfortunately, the case
of two sequence breakpoints, corresponding to so called reversals,
cannot be treated within this framework.

The second application arises in the study of chemical channels. The
chemical channel is a channel model in which symbols are used to describe
molecules, and where the channel permutes the molecules in a queue
using adjacent transpositions%
\footnote{Usually, the channel is initialized by a molecule that may appear
in the queue as well.%
}. In information theory, the standard chemical channel model assumes
that there are only two molecules, and that the channel has only two
states - hence the use of adjacent transpositions. If all the molecules
are different, and the channel is allowed to output molecules with
time-varying probabilities, one arrives at a channel model for which
the output is a cost-constrained permutation of the input. Finding
the minimum cost sequence decomposition therefore represents an important
step in the maximum likelihood decoding algorithm fort the channel.

The third application is concerned with flash memories and rank permutation
coding (see \cite{Bruck} and \cite{5485013}). In this case, one
is also interested in sorting permutations using adjacent transpositions
and computing the Kendall distance between permutations \cite{chadwick69}.
If one considers more precise charge leakage models for memory cells,
the costs of adjacent transpositions become non-uniform. This can
easily be captured by a transposition cost model in which non-adjacent
transpositions have unbounded cost, while the costs of adjacent transpositions
are unrestricted. Hence, the proposed decomposition algorithms can
be used as part of general soft-information rank modulation decoders. 

Our findings are organized as follows. Section II introduces the notation
followed in the remainder of the paper, as well as relevant definitions.
Sections III and IV contain the main results of our study: a three-stage
polynomial-time approximation algorithm for general cost-constrained
sorting of permutations, an exact polynomial-time algorithm for sorting
with metric-path costs, as well as a complexity analysis of the described
techniques. Section V contains the concluding remarks.

\section{Notation and Definitions\label{sec:Notation-and-Definitions}}

A permutation $\pi$ of $\{1,2,\cdots,n\}$ is a bijection from $\{1,2,\cdots,n\}$
to itself. The set of permutations of $\{1,2,\cdots,n\}$ is denoted
by $\mathbb{S}_{n}$, and is called the symmetric group on $\{1,2,\cdots,n\}$.
A permutation can be represented in several ways. In the two-line
notation, the domain is written on top, and its image below. The one-line
representation is the second row of the two-line representation. A
permutation may also be represented as the set of elements and their
images. 

For example, one can write a permutation $\pi$ as $\pi\left(1\right)=3,\pi\left(2\right)=1,\pi\left(3\right)=2,\pi\left(4\right)=5,\pi\left(5\right)=4,$
or more succinctly as $\pi=31254$, or in the two-line notation as
\[
\pi=\left(\begin{array}{ccccc}
1 & 2 & 3 & 4 & 5\\
3 & 1 & 2 & 5 & 4\end{array}\right).\]
Yet another way of writing a permutation is via a set of mappings,
for example $\pi=\left\{ 1\rightarrow3,2\rightarrow1,3\rightarrow2,4\rightarrow5,5\rightarrow4\right\} .$
It will be helpful to think of a permutation as a mapping from \emph{positions}
to \emph{objects}. For example, $\pi\left(1\right)=3$ means object
3 occupies position 1. Alternatively, we can also say that element
1 is a predecessor of element 3. If not otherwise stated, the word
\emph{predecessor} will be henceforth used in this context.

The product $\pi_{2}\pi_{1}$ of two permutations $\pi_{1}$ and $\pi_{2}$
is the permutation obtained by first applying $\pi_{1}$ and then
$\pi_{2}$ to $\{1,2,\cdots,n\}$, i.e., the product represents the
composition of $\pi_{1}$ and $\pi_{2}$. 

The \emph{functional digraph}\textbf{ }of a function $f:\{1,2,\cdots,n\}\rightarrow\{1,2,\cdots,n\}$,
denoted by $\mathcal{G}\left(f\right)$, is a directed graph with
vertex set $\{1,2,\cdots,n\}$ and an edge from $i$ to $f\left(i\right)$
for each $i\in\{1,2,\cdots,n\}$. We use the words vertex and element
interchangeably. For a permutation $\pi$ of $\{1,2,\cdots,n\}$,
$\mathcal{G}\left(\pi\right)$ is a collection of disjoint \emph{cycles,}
since the in-degree and out-degree of each vertex is exactly one.
Each cycle can be written as a $k-$tuple $\sigma=\left(a_{1}a_{2}\cdots a_{k}\right)$,
where $k$ is the length of the cycle and $a_{i+1}=\sigma\left(a_{i}\right)$.
For each cycle of length $k$, the indices are evaluated modulo $k$,
so that $a_{k+1}$ equals $a_{1}$. A planar embedding of $\mathcal{G}\left(\pi\right)$
can be obtained by placing vertices of each of the disjoint cycles
on disjoint circles\emph{.} We hence reserve the symbol $\sigma$
for single cycles, and $\pi$ for multiple cycle permutations. 

We use $\mathcal{G}\left(\pi\right)$ to refer to the planar embedding
of the functional digraph of $\pi$ on circles, as well as the functional
digraph of $\pi$. As a convention, we do not explicitly indicate
the direction of edges on the circle. Instead, we assume a clockwise
direction and treat $\mathcal{G}\left(\pi\right)$ as a non-directional
graph, unless otherwise stated. 

A cycle of length two is called a \emph{transposition}. A \emph{transposition
decomposition}\textbf{ $\tau$} (or simply a decomposition) of a permutation
$\pi$ is a sequence $t_{m}\cdots t_{1}$ of transpositions $t_{i}$
whose product is $\pi$. Note that the transpositions are applied
from right to left. A \emph{sorting}\textbf{ }$s$ of a permutation
$\pi$ is a sequence of transpositions that transform $\pi$ into
$\imath$, where $\imath$ denotes the identity element of $\mathbb{S}_{n}$.
In other words, $s\pi=\imath$. Note that a decomposition $\tau$
in reverse order equals a sorting $s$ of the same permutation. 

The cycle representation of a permutation is the list of its cycles.
For example, the cycle representation of $31254$ is $\left(132\right)\left(45\right)$.
Cycles of length one are usually omitted. The product of non-disjoint
cycles is interpreted as a product of permutations. As an illustration,
$\left(124\right)\left(213\right)=\left(\left(124\right)\left(3\right)\right)\left(\left(213\right)\left(4\right)\right)=\left(2\right)\left(134\right)$. 

A permutation $\pi$ is said to be odd (even) if the number of pairs
$a,b\in\{1,2,\cdots,n\}$ such that $a<b$ and $\pi\left(a\right)>\pi\left(b\right)$
is odd (even). If a permutation is odd (even), then the number of
transpositions in any of its decompositions is also odd (even).

The following definitions regarding graphs $G=\left(V,E\right)$ will
be\emph{ }used throughout the paper. An edge with endpoints $u$ and
$v$ is denoted by $\left(uv\right)\in E$. A graph is said to be
planar if it can be embedded in the plane without intersecting edges.
The subgraph of $G$ induced by the vertices in the set $S\subset V$
is denoted by $G\left[S\right]$. The degree of a vertex $v$ in $G$
is denoted by $\deg_{G}\left(v\right)$ or, if there is no ambiguity,
by $\deg\left(v\right)$. Deletion of an edge $e$ from a graph $G$
is denoted by $G-e$ and deletion of a vertex $v$ and its adjacent
edges from $G$ is denoted by $G-v$. The same notions can be defined
for multigraphs - graphs in which there may exist multiple edges between
two vertices. 

We say that an edge $e$ in $G$ is a cut edge for two vertices $a$
and $b$, denoted by $\cutedge ab$, if in $G-e$ there exists no
path between $a$ and $b$. The well known Menger's theorem\cite{doug_west_comb_math}
asserts that the minimum number of edges one needs to delete from
$G$ to disconnect $a$ from $b$ is also the maximum number of pairwise
edge-disjoint paths between $a$ and $b.$ This theorem holds for
multigraphs as well.

Let $\mathcal{T}\left(\tau\right)$ be a (multi)graph with vertex
set $\{1,2,\cdots,n\}$ and edges $(a_{i}b_{i})$ for each transposition
$t_{i}=\left(a_{i}b_{i}\right)$ of $\tau$. We use the words transposition
and edge interchangeably. The embedding of $\mathcal{T}\left(\tau\right)$
with vertex set $\{1,2,\cdots,n\}$ into $\mathcal{G}\left(\pi\right)$
is also denoted by $\mathcal{T}\left(\tau\right)$. In the derivations
to follow, we make frequent use of the spanning trees of the (multi)graphs
$\mathcal{T}\left(\tau\right),\mathcal{G}\left(\pi\right)$ and $\mathcal{G}\left(\pi\right)\cup\mathcal{T}\left(\tau\right)$.
A spanning tree is a standard notion in graph theory: it is a tree
that contains all vertices of the underlying (multi)graph.

We are concerned with the following problem: given a non-negative
cost function $\varphi$ on the set of transpositions, the cost of
a transposition decomposition is defined as the sum of costs of its
transpositions. The task is to find an efficient algorithm for generating
the Minimum Cost Transposition Decomposition (MCD) of a permutation
$\pi\in\mathbb{S}_{n}$ . The cost of the MCD of a permutation $\pi$
under cost function $\varphi$ is denoted by $M_{\varphi}\left(\pi\right)$.
If there is no ambiguity, the subscript is omitted.

For a non-negative cost function $\varphi$, let $\mathcal{K}\left(\varphi\right)$
be the undirected complete graph in which the cost of each edge $(ab)$
equals $\varphi\left(a,b\right)$. The cost of a graph $ $$G\subseteq\mathcal{K}\left(\varphi\right)$
is the sum of the costs of its edges, \[
\cost\left(G\right)=\sum_{\left(ab\right)\in G}\varphi\left(a,b\right).\]
The shortest path, i.e., the path with minimum cost, between $i$
and $j$ in $\mathcal{K}\left(\varphi\right)$ is denoted by $p^{*}\left(i,j\right)$.

The following definitions pertaining to cost functions are useful
in our analysis. A cost function $\varphi$ is a metric if for $a,b,c\in\{1,2,\cdots,n\}$\[
\varphi\left(a,c\right)\le\varphi\left(a,b\right)+\varphi\left(b,c\right).\]
A cost function $\varphi$ is a\textbf{ }\emph{metric-path}\textbf{
}cost if it is defined in terms of a weighted path, denoted by $\Theta_{s}$.
The weights of edges $\left(uv\right)$ in $\Theta_{s}$ are equal
to $\varphi\left(u,v\right)$, and the cost of any transposition $\left(ij\right)$
equals \[
\varphi\left(i,j\right)=\sum_{t=1}^{l}\varphi\left(c_{t},c_{t+1}\right),\]
where $c_{1}\cdots c_{l}c_{l+1}$, $c_{1}=i$, $c_{l+1}=j$, represents
the unique path between $i$ and $j$ in $\Theta_{s}$. The path $\Theta_{s}$
is called the \emph{defining path}\textbf{ }of $\varphi$. A cost
function $\varphi$ is an\textbf{ }\emph{extended-metric-path}\textbf{
}cost function if for a defining path $\Theta_{s}$, $\varphi\left(i,j\right)$
is finite only for the edges $\left(ij\right)$ of the defining path,
and unbounded otherwise. 

\global\long\def\htrans#1#2#3{\left(#3,\left(#1\rightarrow#2\right)\right)}

Applying a transposition $\left(ab\right)$ to a permutation $\pi$
is equivalent to exchanging the predecessors of $a$ and $b$ in $\mathcal{G}\left(\pi\right)$.
We define a generalization of the notion of a transposition, termed
\emph{h-transposition}, where the predecessor of $a$ can be changed
independently of the predecessor of $b$. For example, let $a,b,c,d\in\{1,2,\cdots,n\}$
and let $\pi\left(c\right)=a$ and $\pi\left(d\right)=b$. Let $\pi'=\htrans abc\pi$,
where we used $\htrans abc$ to denote an h-transposition. This h-transposition
takes $c$, the predecessor of $a$, to $b$, without modifying the
predecessor of $b$. That is, we have a mapping in which $\pi'\left(c\right)=\pi'\left(d\right)=b$,
and $a$ has no predecessor. Note that $\pi'$ is no longer a bijection,
and several elements may be mapped to one element. A transposition
represents the product of a pair of h-transpositions, as in \[
\left(ab\right)\pi=\htrans ab{\pi^{-1}\left(a\right)}\htrans ba{\pi^{-1}\left(b\right)}\pi.\]

An \emph{h-decomposition}\textbf{ }$h$ of a permutation $\pi$ is
a sequence of h-transpositions such that $h\imath=\pi$. Similar to
transpositions, a cost $\psi\left(a,b\right)\ge0$ can be assigned
to h-transpositions $\htrans abc$, where $c$ is the predecessor
of $a$. Note that the cost $\psi$ is not dependent on $c$. We say
that the transposition cost $\varphi$ and the h-transposition cost
$\psi$ are \emph{consistent }if for all transpositions $\left(ab\right)$
it holds that $\varphi\left(a,b\right)=\psi\left(a,b\right)+\psi\left(b,a\right)$. 

For a permutation $\pi$ and a transposition $\left(ab\right)$, it
can be easily verified that $\left(ab\right)\pi$ consist of one more
(or one less) cycle than $\pi$ if and only if $a$ and $b$ are in
the same cycle (in different cycles). Since the identity permutation
has $n$ cycles, a \emph{Minimum Length Transposition Decomposition
(MLD)} of $\pi$ has length $n-\ell$, where $\ell$ denotes the number
of cycles in $\pi$. The minimum cost of an MLD of $\pi$, with respect
to cost function $\varphi$, is denoted by $L_{\varphi}\left(\pi\right)$.
For example, $\left(132\right)\left(45\right)=\left(45\right)\left(23\right)\left(12\right)$
is decomposed into three transpositions. In particular, if $\pi$
is a single cycle, then the MLD of the cycle has length $n-1$. A
cycle of length $k$ has $k^{k-2}$ MLDs \cite{NilsMeier03011976}.
An MCD is not necessarily an MLD, as illustrated by the following
example. 
\begin{example}
\label{exa:navin}Consider the cycle $\sigma=(1\cdots5)$ with $\varphi\left(i,i+1\right)=3$
and $\varphi\left(i,i+2\right)=1$. It is easy to verify that $\left(14\right)(13)(35)(24)(14)(13)$
is an MCD of $\sigma$ with cost six, i.e., $M\left(\sigma\right)=6$.
However, as we shall see later, the cost of a minimum cost MLD is
eight, i.e., $L\left(\sigma\right)=8$. One such MLD is $\left(14\right)\left(23\right)\left(13\right)\left(45\right)$
\cite{navin_personal_2010}.\queede
\end{example}
Our approach to finding the minimum cost decomposition of a permutation
consists of three stages:
\begin{enumerate}
\item First, we find the minimum cost decomposition for each individual
transposition. In particular, we show that the minimum cost decomposition
of a transposition can be obtained by recursively substituting transpositions
with triples of transpositions. This step is superfluous for the case
when the cost function is a metric.
\item In the second step, we consider cycles only and assume that each transposition
cost is optimized. Cycles have the simplest structure among all permutations,
and furthermore, each permutation is a collection of cycles. Hence,
several approximation algorithms operate on individual cycles and
combine their decompositions. As part of this line of results, we
describe how to find the minimum cost MLD and show that its cost is
not more than a constant factor higher than that of the corresponding
MCD. We also present a particularly simple-to-implement class of decompositions
whose costs lie between the cost of a minimum MLD and a constant multiple
of the cost of an MCD.
\item We generalize the results obtained for single cycles to permutations
with multiple cycles. 
\end{enumerate}

\section{Optimizing Individual Transposition\label{sec:Optimizing-Individual-Transposition}}

Let $\tau$ be a transposition decomposition and let $\left(ab\right)$
be a transposition in $\tau$. Since a transposition is an odd permutation,
it may only be written as the composition of an odd number of transpositions.
For example, \begin{equation}
\left(ab\right)=\left(ac\right)\left(bc\right)\left(ac\right),\label{eq:trans_decompose}\end{equation}
where $c\in\{1,2,\cdots,n\}$ and $c\neq a,b$. It is straightforward
to see that any decomposition of a transposition of length three must
be of the form \eqref{eq:trans_decompose}, with a possible reversal
of the roles of the elements $a$ and $b$.

If $\varphi\left(ab\right)>2\varphi\left(ac\right)+\varphi\left(bc\right),$
then replacing $\left(ab\right)$ by $\left(ac\right)\left(bc\right)\left(ac\right)$
reduces the overall cost of $\left(ab\right)$. Thus, the first step
of our decomposition algorithm is to find the optimal cost of each
transposition. As will be shown, it is straightforward to develop
an algorithm for finding \emph{minimum cost decompositions of transpositions
of the form }\eqref{eq:trans_decompose}. One such algorithm -- Alg.
\ref{alg:opt-trans} -- performs a simple search on the ordered set
of transpositions in order to check if their product, of the form
of \eqref{eq:trans_decompose}, yields a decomposition of lower cost
for some transposition. It then updates the costs of transpositions
and performs a new search for decompositions of length three that
may reduce some transposition cost. 

The optimized costs produced by the algorithm are denoted by $\varphi^{*}$.
Note that $\varphi^{*}(a,b)\leq2\varphi^{*}(a,x)+\varphi^{*}(b,x)$,
for any $x\neq a,b.$ Although an optimal decomposition of the form
produced by Alg. \ref{alg:opt-trans}is not guaranteed to produce
the overall minimum cost decomposition of any transposition, we show
that this is indeed the case after the expositions associated with
Alg. \ref{alg:opt-trans}.

Observe that if the cost function is such that \begin{equation}
\varphi\left(b,c\right)+2\varphi\left(a,c\right)\geq\varphi\left(a,b\right),\quad a,b,c\in\{1,2,\cdots,n\},\label{eq:V-ineq}\end{equation}
as in Example 1, Alg. \ref{alg:opt-trans} is redundant and can be
omitted when computing the MCD. In particular, if the cost function
is a metric, then Alg. \ref{alg:opt-trans} is not needed.

The input to the algorithm Alg. \ref{alg:opt-trans} is an ordered
list $\Omega$ of transpositions and their costs. Each row of $\Omega$
corresponds to one transposition and is of the form $\left[\left(ab\right)|\varphi\left(a,b\right)\right]$.
Sorting of $\Omega$ means reordering its rows so that transpositions
are sorted in increasing order of their costs. The output of the algorithm
is a list with the same format, but with minimized costs for each
transposition.

\begin{algorithm}
\caption{\textsc{Optimize-Transposition-Costs$\left(\Omega\right)$}}
\label{alg:opt-trans}

\begin{algorithmic}[1]
\State Input: $\Omega$ (the list of transpositions and their cost)
\State Sort $\Omega$
\For {$i \leftarrow 2\upto \left|\Omega\right|$}
   \State $(a_1 b_1)\leftarrow  \Omega(i)$
   \State $\phi_1 \leftarrow  \varphi {\left(a_1,b_1\right)}$
   \For {$j=1 \upto i-1$}
      \State $(a_2 b_2)\leftarrow  \Omega(j)$
      \State $\phi_2 \leftarrow \varphi {\left(a_2,b_2\right)}$
	  \If {$\{a_1 b_1\} \cap \{a_2 b_2\}\neq\emptyset$}
     	\State $a_{com} \leftarrow  \{a_1,b_1\}\cap\{a_2,b_2\}$
 		\State $\{a_3,b_3\}\leftarrow \{a_1,a_2,b_1,b_2\}-\{a_{com}\}$
 		\If {$\phi_{1}+2\phi_{2}<\varphi {\left(a_3,b_3\right)}$}
				\State update $\varphi {\left(a_3,b_3\right)}$ in $\Omega$
         \EndIf
      \EndIf
	\EndFor
   \State Sort $\Omega$
\EndFor
\end{algorithmic}
\end{algorithm}

\begin{lem}
\label{lem:opt-trans} Alg. \ref{alg:opt-trans} optimizes the costs
$\varphi$ of all transpositions with respect to the triple transposition
decomposition. \end{lem}
\begin{IEEEproof}
Let $\Omega_{i}$ be the list $\Omega$ at the beginning of iteration
$i$, obtained immediately before executing line 4 of Alg. \ref{alg:opt-trans}.
We prove, by induction, that transpositions in $\Omega_{i}(1\upto i)$
have minimum triple decomposition costs that do not change in subsequent
iterations of the algorithm, and that the transpositions in $\Omega_{i}(i+1\upto\left|\Omega\right|)$
cannot be written as a product of transpositions exclusively in $\Omega_{i}(1:i-1)$
that have smaller overall cost.

The claim is obviously true for $i=2$.

Assume that the claim holds for $i$. Let $t_{1}=\Omega_{i}(i)$ and
consider $s\in\Omega_{i}(i+1:\left|\Omega\right|)$. By the induction
assumption, $s$ cannot be written as a product of transpositions
exclusively in $\Omega_{i}(1:i-1)$ having smaller overall cost. Thus,
the cost of $s$ may be reduced only if one can write $s$ as $t_{2}t_{1}t_{2}$,
where $t_{2}\in\Omega(1\upto i-1)$. The list $\Omega_{i+1}$ is obtained
after considering all such transpositions, updating $\varphi$ and
sorting $\Omega_{i}$. The transposition of minimum cost in $\Omega_{i+1}(i+1\upto\left|\Omega\right|)$
is $\Omega_{i+1}(i+1)$. Now $\Omega_{i+1}(i+1\upto\left|\Omega\right|)$
cannot be written in terms of transpositions in $\Omega_{i+1}(1:i)$
only, and hence the cost of any transposition in $\Omega_{i+1}(i+1:\left|\Omega\right|)$
cannot be reduced below the cost of $\Omega_{i+1}(i+1)$. Hence, the
cost of $\Omega_{i+1}(i+1)$ is minimized. \end{IEEEproof}
\begin{example}
The left-most list in \eqref{eq:trans_opt_ex} represents the input
$\Omega$ to the algorithm, with transpositions in increasing order
of their costs. The two lists that follow represent updates of $\Omega$
produced by Alg. \ref{alg:opt-trans}. In the first step, the algorithm
considers the transposition $(13)$, for $i=2$, and the transposition
$\left(34\right)$, for $j=1$. Using these transpositions we may
write $\left(34\right)\left(13\right)\left(34\right)=\left(14\right)$.
The initial cost of $\left(14\right)$ is 12 which exceeds $2\varphi\left(3,4\right)+\varphi\left(1,3\right)=8$.
Hence, the list representing $\Omega$ is updated to form the second
list in \eqref{eq:trans_opt_ex}. Next, for $i=3$ and $j=1$, the
algorithm considers $\left(24\right)$ and $\left(34\right)$. Since
$\left(34\right)\left(24\right)\left(34\right)=(23)$, we update the
cost of $\left(23\right)$ from $23$ to $11$ as shown in the third
list in \eqref{eq:trans_opt_ex}. Additional iterations of the algorithm
introduce no further changes in the costs. 
\end{example}
\begin{equation}
\left[\begin{array}{c|c}
\left(34\right) & 2\\
\left(13\right) & 4\\
\left(24\right) & 7\\
\left(14\right) & 12\\
\left(12\right) & 15\\
\left(23\right) & 23\end{array}\right]\rightarrow\left[\begin{array}{c|c}
\left(34\right) & 2\\
\left(13\right) & 4\\
\left(24\right) & 7\\
\mathit{\left(14\right)} & \mathit{8}\\
\left(12\right) & 15\\
\left(23\right) & 23\end{array}\right]\rightarrow\left[\begin{array}{c|c}
\left(34\right) & 2\\
\left(13\right) & 4\\
\left(24\right) & 7\\
\left(14\right) & 8\\
\mathit{\left(23\right)} & \mathit{11}\\
\left(12\right) & 15\end{array}\right]\label{eq:trans_opt_ex}\end{equation}
\queede

Upon executing the algorithm, the cost of each transposition is set
to its minimal value. Only after the last stage of the MCD approximation
algorithm is completed will each transposition be replaced by its
minimal cost decomposition. For each index $i$ the number of operations
performed in the algorithm is $O(\left|\Omega\right|)$. Thus, the
total complexity of the algorithm is $O(\left|\Omega\right|^{2})$.
Since $\left|\Omega\right|$ is at most equal to the number of transpositions,
we have $\left|\Omega\right|={n \choose 2}$. Hence, the complexity
of Alg. \ref{alg:opt-trans} equals $O(n^{4})$.

In the analysis that follows, denote the optimized transposition costs
by the superscript $*$, as in $\varphi^{*}$. 

Since the transposition costs are arbitrary non-negative values, it
is not clear that the minimum cost decomposition of a transposition
is necessarily of the form generated by Alg. \ref{alg:opt-trans}.
This algorithm only guarantees that one can identify the \emph{optimal
sequence of consecutive replacements of transpositions by triples
of transpositions}. Hence, the minimum cost of a transposition $(ab)$
may be smaller than $\varphi^{*}(a,b)$, i.e. there may be decompositions
of length five, seven, or longer, which allow for an even smaller
decomposition cost of a transposition. 

Fortunately, this is not the case: we first prove this claim for decompositions
of length five via exhaustive enumeration and then proceed to prove
the general case via the use of Mengers's theorem for multigraphs\cite{goulden_combinatorial_2004}.
We choose to provide the example of length-five decompositions since
it illustrates the difficulty of proving statements about non-minimal
decompositions of permutations using exhaustive enumeration techniques.
Graphical representations, on the other hand, allow for much more
general and simpler proofs pertaining to non-minimal decompositions
of transpositions.

We start by considering all possible transposition decompositions
of length five, for which the transposition costs are first optimized
via Alg. \ref{alg:opt-trans}. In other words, we investigate if there
exist decompositions of $(ab)$ of length five that have cost smaller
than $\varphi^{*}(a,b)$. Once again, observe that the costs of all
transpositions used in such decompositions are first optimized via
a sequence of triple-transposition decompositions. To reduce the number
of cases, we present the following lemma restricting the possible
configurations in a multigraph corresponding to the decomposition
of a transposition $\left(ab\right)$.
\begin{lem}
Let $\tau$ be a decomposition of a transposition $\left(ab\right)$.
The multigraph $\mathcal{M}=\mathcal{T}\left(\tau\right)$, where
$\tau$ does not contain $\left(ab\right)$, has the following properties:\label{lem:properties-of-T}\end{lem}
\begin{enumerate}
\item Both $a$ and $b$ have degree at least one.\label{enu:deg-a-b-1}
\item The degree of at least one of the vertices $a$ and $b$ is at least
two.\label{enu:deg-a-b-2}
\item Every vertex of $\mathcal{M}=\mathcal{T}\left(\tau\right)$, other
than $a$ and $b$, appears in a closed path (cycle) with no repeated
edges in $\mathcal{M}$. \label{enu:Every-vertex-cycle}\end{enumerate}
\begin{IEEEproof}
The proof follows from the simple observations that~:
\begin{enumerate}
\item In order to swap $a$ and $b$, both $a$ and $b$ must be moved.
\item If both vertices $a$ and $b$ have degree one, then $a$ and $b$
are moved exactly once. This is only possible only if $\left(ab\right)\in\tau.$
\item Let $\tau=t_{m}\cdots t_{2}t_{1}$. Let $t_{i}$ be the transposition
with the smallest index $i$ that includes $x\in V(\mathcal{M})$.
In the permutation $\tau_{1}=t_{i}\cdots t_{2}t_{1}$, $x$ is not
in its original location but rather occupies the position of another
element, say, $y$. As we shall see in the proof of Lemma \eqref{lem:MLD-graph}
and Example \eqref{exa:path-from-a-to-b-(ab)}, this means that there
is a path from $x$ to $y$ in $\mathcal{T}(\tau_{1})$. Similarly,
there must exist a path from $y$ to $x$ in $\mathcal{T}\left(t_{m}\cdots t_{i+2}t_{i+1}\right)$.
Thus there is a closed path with no repeated edges from $x$ to itself
in $\mathcal{M}$.
\end{enumerate}
\end{IEEEproof}
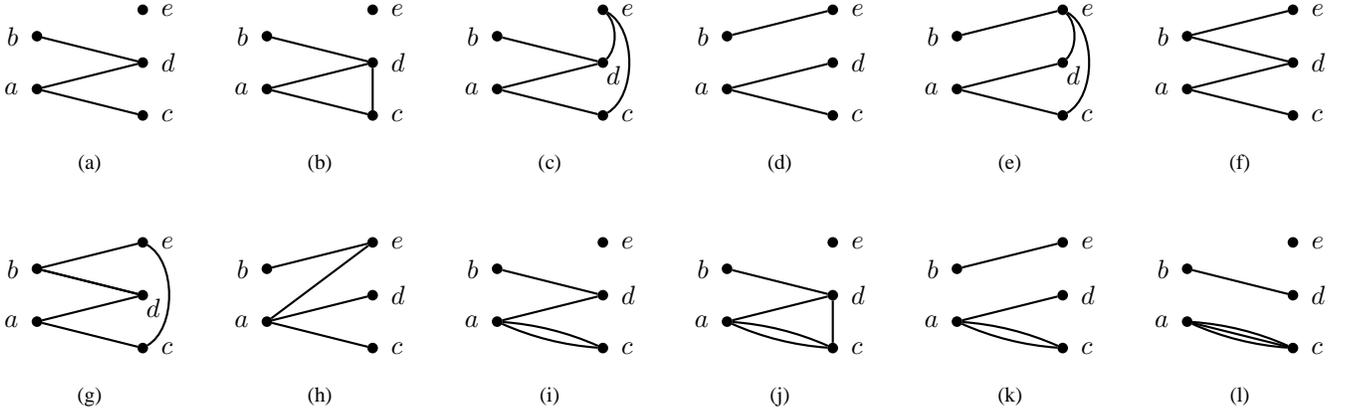
\begin{figure*}
\subfloat[]{
\centering

\psset{unit=20pt}

\begin{pspicture}[showgrid=false](-1,-.5)(3, 2.75)

\cnode*(0,.5){2pt}{a}

\nput{180}{a}{$a$}

\cnode*(0,1.5){2pt}{b}

\nput{180}{b}{$b$}

\cnode*(2,0){2pt}{c}

\nput{0}{c}{$c$}

\cnode*(2,1){2pt}{d}

\nput{0}{d}{$d$}

\cnode*(2,2){2pt}{e}

\nput{0}{e}{$e$}

\ncarc[arcangle=0,linestyle=solid]{a}{c}

\ncarc[arcangle=0,linestyle=solid]{a}{d}

\ncarc[arcangle=0,linestyle=solid]{b}{d}

\end{pspicture}

\label{fig:ab21-1}}
\subfloat[]{
\centering

\psset{unit=20pt}

\begin{pspicture}[showgrid=false](-1,-.5)(3, 2.75)

\cnode*(0,.5){2pt}{a}

\nput{180}{a}{$a$}

\cnode*(0,1.5){2pt}{b}

\nput{180}{b}{$b$}

\cnode*(2,0){2pt}{c}

\nput{0}{c}{$c$}

\cnode*(2,1){2pt}{d}

\nput{0}{d}{$d$}

\cnode*(2,2){2pt}{e}

\nput{0}{e}{$e$}

\ncarc[arcangle=0,linestyle=solid]{a}{c}

\ncarc[arcangle=0,linestyle=solid]{a}{d}

\ncarc[arcangle=0,linestyle=solid]{b}{d}

\ncarc[arcangle=0,linestyle=solid]{c}{d}

\end{pspicture}

\label{fig:ab21-cd}}
\subfloat[]{
\centering

\psset{unit=20pt}

\begin{pspicture}[showgrid=false](-1,-.5)(3, 2.75)

\cnode*(0,.5){2pt}{a}

\nput{180}{a}{$a$}

\cnode*(0,1.5){2pt}{b}

\nput{180}{b}{$b$}

\cnode*(2,0){2pt}{c}

\nput{0}{c}{$c$}

\uput*{.1}[-45](2,1){$d$}

\cnode*(2,1){2pt}{d}

\cnode*(2,2){2pt}{e}

\nput{0}{e}{$e$}

\ncarc[arcangle=0,linestyle=solid]{a}{c}

\ncarc[arcangle=0,linestyle=solid]{a}{d}

\ncarc[arcangle=0,linestyle=solid]{b}{d}

\ncarc[arcangle=60,linestyle=solid]{e}{c}

\ncarc[arcangle=45,linestyle=solid]{e}{d}

\end{pspicture}

\label{fig:ab21-notcd}}
\subfloat[]{
\centering

\psset{unit=20pt}

\begin{pspicture}[showgrid=false](-1,-.5)(3, 2.75)

\cnode*(0,.5){2pt}{a}

\nput{180}{a}{$a$}

\cnode*(0,1.5){2pt}{b}

\nput{180}{b}{$b$}

\cnode*(2,0){2pt}{c}

\nput{0}{c}{$c$}

\cnode*(2,1){2pt}{d}

\nput{0}{d}{$d$}

\cnode*(2,2){2pt}{e}

\nput{0}{e}{$e$}

\ncarc[arcangle=0,linestyle=solid]{a}{c}

\ncarc[arcangle=0,linestyle=solid]{a}{d}

\ncarc[arcangle=0,linestyle=solid]{b}{e}

\end{pspicture}

\label{fig:ab21-be}}
\subfloat[]{
\centering

\psset{unit=20pt}

\begin{pspicture}[showgrid=false](-1,-.5)(3, 2.75)

\cnode*(0,.5){2pt}{a}

\nput{180}{a}{$a$}

\cnode*(0,1.5){2pt}{b}

\nput{180}{b}{$b$}

\cnode*(2,0){2pt}{c}

\nput{0}{c}{$c$}

\uput*{.1}[-45](2,1){$d$}

\cnode*(2,1){2pt}{d}

\cnode*(2,2){2pt}{e}

\nput{0}{e}{$e$}

\ncarc[arcangle=0,linestyle=solid]{a}{c}

\ncarc[arcangle=0,linestyle=solid]{a}{d}

\ncarc[arcangle=0,linestyle=solid]{b}{e}

\ncarc[arcangle=-60,linestyle=solid]{c}{e}

\ncarc[arcangle=-45,linestyle=solid]{d}{e}

\end{pspicture}

\label{fig:ab21-adec}}
\subfloat[]{
\centering

\psset{unit=20pt}

\begin{pspicture}[showgrid=false](-1,-.5)(3, 2.75)

\cnode*(0,.5){2pt}{a}

\nput{180}{a}{$a$}

\cnode*(0,1.5){2pt}{b}

\nput{180}{b}{$b$}

\cnode*(2,0){2pt}{c}

\nput{0}{c}{$c$}

\cnode*(2,1){2pt}{d}

\nput{0}{d}{$d$}

\cnode*(2,2){2pt}{e}

\nput{0}{e}{$e$}

\ncarc[arcangle=0,linestyle=solid]{a}{c}

\ncarc[arcangle=0,linestyle=solid]{a}{d}

\ncarc[arcangle=0,linestyle=solid]{b}{d}

\ncarc[arcangle=0,linestyle=solid]{b}{e}

\end{pspicture}

\label{fig:ab22}}

\subfloat[]{
\centering

\psset{unit=20pt}

\begin{pspicture}[showgrid=false](-1,-.5)(3, 2.75)

\cnode*(0,.5){2pt}{a}

\nput{180}{a}{$a$}

\cnode*(0,1.5){2pt}{b}

\nput{180}{b}{$b$}

\cnode*(2,0){2pt}{c}

\nput{0}{c}{$c$}

\uput*{.1}[-45](2,1){$d$}

\cnode*(2,1){2pt}{d}

\cnode*(2,2){2pt}{e}

\nput{0}{e}{$e$}

\ncarc[arcangle=0,linestyle=solid]{a}{c}

\ncarc[arcangle=0,linestyle=solid]{a}{d}

\ncarc[arcangle=0,linestyle=solid]{b}{d}

\ncarc[arcangle=0,linestyle=solid]{b}{e}

\ncarc[arcangle=0,linestyle=solid]{b}{d}

\ncarc[arcangle=-60,linestyle=solid]{c}{e}

\end{pspicture}

\label{fig:ab22ec}}
\subfloat[]{
\centering

\psset{unit=20pt}

\begin{pspicture}[showgrid=false](-1,-.5)(3, 2.75)

\cnode*(0,.5){2pt}{a}

\nput{180}{a}{$a$}

\cnode*(0,1.5){2pt}{b}

\nput{180}{b}{$b$}

\cnode*(2,0){2pt}{c}

\nput{0}{c}{$c$}

\cnode*(2,1){2pt}{d}

\nput{0}{d}{$d$}

\cnode*(2,2){2pt}{e}

\nput{0}{e}{$e$}

\ncarc[arcangle=0,linestyle=solid]{a}{c}

\ncarc[arcangle=0,linestyle=solid]{a}{d}

\ncarc[arcangle=0,linestyle=solid]{a}{e}

\ncarc[arcangle=0,linestyle=solid]{b}{e}

\end{pspicture}

\label{fig:ab31e}}
\subfloat[]{
\centering

\psset{unit=20pt}

\begin{pspicture}[showgrid=false](-1,-.5)(3, 2.75)

\cnode*(0,.5){2pt}{a}

\nput{180}{a}{$a$}

\cnode*(0,1.5){2pt}{b}

\nput{180}{b}{$b$}

\cnode*(2,0){2pt}{c}

\nput{0}{c}{$c$}

\cnode*(2,1){2pt}{d}

\nput{0}{d}{$d$}

\cnode*(2,2){2pt}{e}

\nput{0}{e}{$e$}

\ncarc[arcangle=10,linestyle=solid]{a}{c}

\ncarc[arcangle=-10,linestyle=solid]{a}{c}

\ncarc[arcangle=0,linestyle=solid]{a}{d}

\ncarc[arcangle=0,linestyle=solid]{b}{d}

\end{pspicture}

\label{fig:ab31d}}
\subfloat[]{
\centering

\psset{unit=20pt}

\begin{pspicture}[showgrid=false](-1,-.5)(3, 2.75)

\cnode*(0,.5){2pt}{a}

\nput{180}{a}{$a$}

\cnode*(0,1.5){2pt}{b}

\nput{180}{b}{$b$}

\cnode*(2,0){2pt}{c}

\nput{0}{c}{$c$}

\cnode*(2,1){2pt}{d}

\nput{0}{d}{$d$}

\cnode*(2,2){2pt}{e}

\nput{0}{e}{$e$}

\ncarc[arcangle=10,linestyle=solid]{a}{c}

\ncarc[arcangle=-10,linestyle=solid]{a}{c}

\ncarc[arcangle=0,linestyle=solid]{a}{d}

\ncarc[arcangle=0,linestyle=solid]{b}{d}

\ncarc[arcangle=0,linestyle=solid]{c}{d}

\end{pspicture}

\label{fig:ab31ddc}}
\subfloat[]{
\centering

\psset{unit=20pt}

\begin{pspicture}[showgrid=false](-1,-.5)(3, 2.75)

\cnode*(0,.5){2pt}{a}

\nput{180}{a}{$a$}

\cnode*(0,1.5){2pt}{b}

\nput{180}{b}{$b$}

\cnode*(2,0){2pt}{c}

\nput{0}{c}{$c$}

\cnode*(2,1){2pt}{d}

\nput{0}{d}{$d$}

\cnode*(2,2){2pt}{e}

\nput{0}{e}{$e$}

\ncarc[arcangle=10,linestyle=solid]{a}{c}

\ncarc[arcangle=-10,linestyle=solid]{a}{c}

\ncarc[arcangle=0,linestyle=solid]{a}{d}

\ncarc[arcangle=0,linestyle=solid]{b}{e}

\end{pspicture}

\label{fig:ab31cc}}
\subfloat[]{
\centering

\psset{unit=20pt}

\begin{pspicture}[showgrid=false](-1,-.5)(3, 2.75)

\cnode*(0,.5){2pt}{a}

\nput{180}{a}{$a$}

\cnode*(0,1.5){2pt}{b}

\nput{180}{b}{$b$}

\cnode*(2,0){2pt}{c}

\nput{0}{c}{$c$}

\cnode*(2,1){2pt}{d}

\nput{0}{d}{$d$}

\cnode*(2,2){2pt}{e}

\nput{0}{e}{$e$}

\ncarc[arcangle=10,linestyle=solid]{a}{c}

\ncarc[arcangle=-10,linestyle=solid]{a}{c}

\ncarc[arcangle=0,linestyle=solid]{a}{c}

\ncarc[arcangle=0,linestyle=solid]{b}{d}

\end{pspicture}

\label{fig:ab31ccc}}

\caption{Possible $G_{1}-$ and $G_{2}-$free configurations for $\mathcal{M}=\mathcal{T}\left(\tau\right)$
when $\tau$ is a five-decomposition of $\left(ab\right)$. Note that
other configurations can be obtained from these by relabeling $a$
and $b$, and relabeling $c,d,$ and $e$ since they are symmetric.}

\end{figure*}

Let $x_{1},x_{2},\cdots,x_{N}$ be vertices included in the decomposition
$\tau$ other than $a$ and $b$. If $E(\mathcal{M})$ denotes the
number of edges in the multigraph $\mathcal{M}=\mathcal{T}(\tau)$,
then \[
2\left|E\left(\mathcal{M}\right)\right|=\sum_{i=1}^{N}\deg\left(x_{i}\right)+\deg\left(a\right)+\deg\left(b\right).\]
From parts \ref{enu:deg-a-b-1} and \ref{enu:deg-a-b-2} of Lemma
\ref{lem:properties-of-T}, note that $\deg\left(a\right)+\deg\left(b\right)\ge3$
and, from part \ref{enu:Every-vertex-cycle}, it holds that $\deg\left(x_{i}\right)\ge2$.
Hence, $2\left|E\left(\mathcal{M}\right)\right|\ge2N+3$, and since
$N$ has to be an integer, \begin{equation}
N\le\lfloor\left|E\left(\mathcal{M}\right)\right|-\frac{3}{2}\rfloor=\left|E\left(\mathcal{M}\right)\right|-2.\label{eq:vertex-bound}\end{equation}
Suppose that $\tau=t_{5}t_{4}t_{3}t_{2}t_{1}$ is the minimum cost
decomposition of $\left(ab\right)$ with cost $\phi$, and that the
cost of the optimal decomposition produced by Alg. \ref{alg:opt-trans}
exceeds $\phi$. Then there is no vertex $x$ such that \[
G_{1}=\left\{ \left(ax\right),\left(ax\right),\left(bx\right)\right\} \]
 is a subset of edges in the multigraph $\mathcal{M}$ since, in that
case, \[
\varphi^{*}\left(a,b\right)\le2\varphi^{*}\left(a,x\right)+\varphi^{*}\left(b,x\right)\le\phi.\]
 Also, there exists no pair of vertices $x,y$ such that \[
G_{2}=\left\{ \left(ax\right),\left(bx\right),\left(ay\right),\left(by\right)\right\} \]
is a subset of edges in the multigraph $\mathcal{M}$. To prove this
claim, suppose that $G_{2}\subseteq E(\mathcal{M})$. Without loss
of generality, assume that \[
\varphi^{*}\left(a,x\right)+\varphi^{*}\left(b,x\right)\le\varphi^{*}\left(a,y\right)+\varphi^{*}\left(b,y\right).\]
Then,\begin{align*}
\varphi^{*}\left(a,b\right) & \le2\varphi^{*}\left(a,x\right)+\varphi^{*}\left(b,x\right)\\
 & \le2\varphi^{*}\left(a,x\right)+2\varphi^{*}\left(b,x\right)\\
 & \le\cost\left(G_{2}\right)\\
 & \le\phi.\end{align*}

Hence, any decomposition of length five that contains $G_{2}$ must
have cost at least $\varphi^{*}\left(a,b\right)$.

For any five-decomposition $\tau$, we have $\left|E\left(\mathcal{M}\right)\right|=5$
and, thus, $N\leq3$. We consider all five-decompositions of $\left(ab\right)$
such that $\mathcal{M}$ is $G_{1}-$free and $G_{2}-$free, and which
contain at most five vertices in $\mathcal{M}$. Assume that the three
extra vertices, in addition to $a$ and $b$, are $c$, $d$, and
$e$. We now show that for each decomposition of length five, there
exists a decomposition obtained via Alg. \ref{alg:opt-trans} with
cost at most $\phi$, denoted by either $\mu$ or $\mu'$. The following
scenarios are possible.
\begin{enumerate}
\item Suppose that $\deg\left(a\right)=2$ and $\deg\left(b\right)=1$.
Furthermore, suppose that there exist a vertex that is adjacent to
both $a$ and $b$ in $\mathcal{M}$. Without loss of generality,
assume that $\left(ad\right)\in\mathcal{M},\left(bd\right)\in\mathcal{M}$
(Figure \ref{fig:ab21-1}). We consider two cases, depending on the
existence of the edge $\left(cd\right)$ in $\mathcal{M}$.\\
First, assume that $\left(cd\right)\in\mathcal{M}$ (Figure \ref{fig:ab21-cd}).
If $\varphi^{*}\left(a,c\right)+\varphi^{*}\left(c,d\right)\le\varphi^{*}\left(a,d\right)$,
then the decomposition \begin{align*}
\mu & =\left(ac\right)\left(cd\right)\left(bd\right)\left(cd\right)\left(ac\right)\end{align*}
has cost at most $\phi$. Note that $\mu$ can be obtained from Alg.
\ref{alg:opt-trans}, since \begin{align*}
\mu & =\left(ac\right)\left(bc\right)\left(ac\right)=\left(ab\right).\end{align*}
On the other hand, if $\varphi^{*}\left(a,c\right)+\varphi^{*}\left(c,d\right)>\varphi^{*}\left(a,d\right)$,
then the decomposition $\mu'=\left(ad\right)\left(bd\right)\left(ad\right)$
has cost at most $\phi$.\\
Next assume that $\left(cd\right)\notin\mathcal{M}$. Since both
$c$ and $d$ each must lie on a cycle, the only possible decompositions
of $(ab)$ are shown in Figure \ref{fig:ab21-notcd}. Now, if $\varphi^{*}\left(ad\right)\le\varphi^{*}\left(d,e\right)+\varphi^{*}\left(e,c\right)+\varphi^{*}\left(a,c\right)$,
then the decomposition \[
\mu=\left(ad\right)\left(bd\right)\left(ad\right)\]
 has cost at most $\phi$. On the other hand, if $\varphi^{*}\left(ad\right)>\varphi^{*}\left(d,e\right)+\varphi^{*}\left(e,c\right)+\varphi^{*}\left(a,c\right)$,
then the decomposition \begin{equation}
\mu'=\left(ac\right)\left(ec\right)\left(ed\right)\left(bd\right)\left(ed\right)\left(ce\right)\left(ac\right)\label{eq:mu-ac-ec-be}\end{equation}
has cost at most $\phi$. Note that $\mu'$ can be obtained from Alg.
\ref{alg:opt-trans}, since \begin{align*}
\mu' & =\left(ac\right)\left(ec\right)\left(be\right)\left(ec\right)\left(ac\right)\\
 & =\left(ac\right)\left(cb\right)\left(ac\right)\\
 & =\left(ab\right).\end{align*}

\item Suppose that $\deg\left(a\right)=2$ and $\deg\left(b\right)=1$,
but that there is no vertex adjacent to both $a$ and $b$. Without
loss of generality, assume $c$ and $d$ are adjacent to $a$ and
$e$ is adjacent to $b$ (Figure \ref{fig:ab21-be}). Since $c,d,$
and $e$ each must lie on a cycle, one must include two more edges
in the graph, as shown in Figure \ref{fig:ab21-adec}. Since $d$
and $c$ have a symmetric role in the decomposition, we may without
loss of generality, assume that $\varphi^{*}\left(a,c\right)+\varphi^{*}\left(c,e\right)\le\varphi^{*}\left(a,d\right)+\varphi^{*}\left(d,e\right)$.
Let $\mu$ be equal to \[
\mu=\left(ac\right)\left(ce\right)\left(eb\right)\left(ce\right)\left(ac\right).\]
Similarly to \eqref{eq:mu-ac-ec-be}, it is easy to see that the cost
of $\mu$ is at most $\phi$ and that it can be obtained from Alg.
\eqref{alg:opt-trans}.
\item Assume that $\deg\left(a\right)=\deg\left(b\right)=2$ (Figure \ref{fig:ab22}).
Since $e$ and $c$ must lie on a cycle, the fifth transposition in
the decomposition must be $\left(ec\right)$ (Figure \ref{fig:ab22ec}).
If $\varphi^{*}\left(a,d\right)+\varphi^{*}\left(b,d\right)\le\varphi^{*}\left(b,e\right)+\varphi^{*}\left(e,c\right)+\varphi^{*}\left(c,a\right)$,
then the decomposition \[
\mu=\left(ad\right)\left(bd\right)\left(ad\right)\]
 has cost at most $\phi$. Otherwise, if $\varphi^{*}\left(a,d\right)+\varphi^{*}\left(b,d\right)>\varphi^{*}\left(b,e\right)+\varphi^{*}\left(e,c\right)+\varphi^{*}\left(c,a\right)$,
the decomposition\[
\mu'=\left(ac\right)\left(ec\right)\left(be\right)\left(ce\right)\left(ac\right)\]
has cost at most $\phi$. Note that both $\mu$ and $\mu'$ represent
decompositions of a form optimized over by Alg. \ref{alg:opt-trans}.
\item Suppose that $\deg\left(a\right)=3$, $\deg\left(b\right)=1$, and
that all edges adjacent to $a$ and $b$ are simple (not repeated).
Without loss of generality, assume that $e$ is adjacent to both $a$
and $b$ (Figure \ref{fig:ab31e}). One edge must complete cycles
that include $c,d,$ and $e$. Since creating such cycles with one
edge is impossible, this configuration is impossible.
\item Suppose that $\deg\left(a\right)=3$, $\deg\left(b\right)=1$, one
edge adjacent to $a$ appears twice, and there is a vertex adjacent
to both $a$ and $b$. Without loss of generality, assume that this
vertex is $d$ (Figure \ref{fig:ab31d}). Since $d$ must be in a
cycle, it must be adjacent to the {}``last edge'', i.e., the fifth
transposition. If the last edge is $\left(ed\right)$, then one more
edge is needed to create a cycle passing through $e$. Thus, the last
edge cannot be $\left(ed\right)$. The only other choice is $\left(cd\right)$
(Figure \ref{fig:ab31ddc}). Now, if $\varphi^{*}\left(a,d\right)\ge\varphi^{*}\left(c,d\right)$,
then the decomposition \[
\mu=\left(ac\right)\left(cd\right)\left(bd\right)\left(cd\right)\left(ac\right)\]
has cost at most $\phi$. Otherwise, if $\varphi^{*}\left(a,d\right)<\varphi^{*}\left(c,d\right)$,
the decomposition \[
\mu'=\left(ad\right)\left(bd\right)\left(ad\right)\]
has cost at most $\phi$.
\item Suppose that $\deg\left(a\right)=3,$ $\deg\left(b\right)=1$, and
no vertex is adjacent to both $a$ and $b$. The two possible cases
are shown in Figures \ref{fig:ab31cc} and \ref{fig:ab31ccc}. Since
one edge cannot create all the necessary cycles, both configurations
are impossible.
\end{enumerate}
\begin{figure*}
\subfloat[The edge $\left(x_{1}y_{1}\right)$ is the only cut-edge]{
\centering

\psset{unit=25pt}

\begin{pspicture}[showgrid=false](-.5,-1.25)(8.25,1.25)

\cnode*(0,0){2pt}{a}

\nput{-180}{a}{$a$}

\nput{0}{a}{$\cdots$}

\psellipse(1.5,0)(1.5,1)

\cnode*(1,0){2pt}{n1}

\cnode*(2,0){2pt}{n2}

\cnode*(1.5,-.25){2pt}{n3}

\cnode*(2,-.5){2pt}{n4}

\cnode*(1,.5){2pt}{n5}

\cnode*(1.5,.25){2pt}{n6}

\ncarc[arcangle=0,linestyle=solid]{n1}{n2}

\ncarc[arcangle=0,linestyle=solid]{n3}{n5}

\ncarc[arcangle=0,linestyle=solid]{n4}{n6}

\psellipse(5.5,0)(1.5,1)

\cnode*(5,0){2pt}{n1}

\cnode*(6,0){2pt}{n2}

\cnode*(5.5,-.25){2pt}{n3}

\cnode*(6,-.5){2pt}{n4}

\cnode*(5,.5){2pt}{n5}

\cnode*(5.5,.25){2pt}{n6}

\ncarc[arcangle=0,linestyle=solid]{n1}{n2}

\ncarc[arcangle=0,linestyle=solid]{n3}{n5}

\ncarc[arcangle=0,linestyle=solid]{n4}{n6}

\cnode*(3,0){2pt}{x1}

\nput{-75}{x1}{$x_1$}

\nput{-180}{x1}{$\cdots$}

\cnode*(4,0){2pt}{y1}

\nput{-110}{y1}{$y_1$}

\ncarc[arcangle=0,linestyle=solid]{y1}{x1}

\psellipse(5.5,0)(1.5,1)

\cnode*(7,0){2pt}{b}

\nput{0}{b}{$b$}

\nput{-180}{b}{$\cdots$}

\nput{0}{y1}{$\cdots$}

\end{pspicture}

\label{fig:one-cut-edge}}
\subfloat[Edges $\left(x_{1}y_{1}\right)$ and $\left(x_{2}y_{2}\right)$ are the cut-edges]{
\centering

\psset{unit=25pt}

\begin{pspicture}[showgrid=false](-.5,-1.25)(8.25,1.25)

\cnode*(0,0){2pt}{a}

\nput{-180}{a}{$a$}

\nput{0}{a}{$\cdots$}

\cnode*(3,0){2pt}{x1}

\nput{-75}{x1}{$x_1$}

\nput{-180}{x1}{$\cdots$}

\cnode*(4,0){2pt}{y1}

\nput{-110}{y1}{$y_1$}

\nput{0}{y1}{$\cdots$}

\ncarc[arcangle=0,linestyle=solid]{y1}{x1}

\cnode*(7,0){2pt}{x2}

\nput{-75}{x2}{$x_2$}

\nput{-180}{x2}{$\cdots$}

\cnode*(8,0){2pt}{y2}

\nput{-110}{y2}{$y_2$}

\nput{0}{y2}{$\cdots$}

\ncarc[arcangle=0,linestyle=solid]{y2}{x2}

\cnode*(11,0){2pt}{b}

\nput{0}{b}{$b$}

\nput{-180}{b}{$\cdots$}

\psellipse(1.5,0)(1.5,1)

\cnode*(1,0){2pt}{n1}

\cnode*(2,0){2pt}{n2}

\cnode*(1.5,-.25){2pt}{n3}

\cnode*(2,-.5){2pt}{n4}

\cnode*(1,.5){2pt}{n5}

\cnode*(1.5,.25){2pt}{n6}

\ncarc[arcangle=0,linestyle=solid]{n1}{n2}

\ncarc[arcangle=0,linestyle=solid]{n3}{n5}

\ncarc[arcangle=0,linestyle=solid]{n4}{n6}

\psellipse(5.5,0)(1.5,1)

\cnode*(5,0){2pt}{n1}

\cnode*(6,0){2pt}{n2}

\cnode*(5.5,-.25){2pt}{n3}

\cnode*(6,-.5){2pt}{n4}

\cnode*(5,.5){2pt}{n5}

\cnode*(5.5,.25){2pt}{n6}

\ncarc[arcangle=0,linestyle=solid]{n1}{n2}

\ncarc[arcangle=0,linestyle=solid]{n3}{n5}

\ncarc[arcangle=0,linestyle=solid]{n4}{n6}

\psellipse(9.5,0)(1.5,1)

\cnode*(9,0){2pt}{n1}

\cnode*(10,0){2pt}{n2}

\cnode*(9.5,-.25){2pt}{n3}

\cnode*(10,-.5){2pt}{n4}

\cnode*(9,.5){2pt}{n5}

\cnode*(9.5,.25){2pt}{n6}

\ncarc[arcangle=0,linestyle=solid]{n1}{n2}

\ncarc[arcangle=0,linestyle=solid]{n3}{n5}

\ncarc[arcangle=0,linestyle=solid]{n4}{n6}

\end{pspicture}

\label{fig:two-cut-edge}}
\caption{ }
\end{figure*}

Next, we state a general theorem pertaining to the optimality of Alg.
\ref{alg:opt-trans}.
\begin{thm}
\label{thm:correctness_of_alg1}The minimum cost decompositions of
all transpositions are generated by Alg. \ref{alg:opt-trans}.\end{thm}
\begin{IEEEproof}
The proof proceeds in two steps. First, we show that the multigraph
$\mathcal{M}$ for a transposition $(ab)$ cannot have more than one
$\cutedge ab$. If $\mathcal{M}$ has no $\cutedge ab$, then there
exist at least two edge-disjoint paths between $a$ and $b$ in $\mathcal{M}$.
This claim follows by invoking Menger's theorem. The costs of the
paths can be combined, leading to a cost of the form induced by a
transposition decomposition optimized via \eqref{eq:trans_decompose}.
If the multigraph has exactly one $\cutedge ab$, this case can be
reduced to the case of no $\cutedge ab$. This completes the proof.

Before proving the impossibility of the existence of more than one
$\cutedge ab$, we explain how a $\cutedge ab$ imposes a certain
structure in the decomposition of $\left(ab\right)$. Consider the
decomposition $t_{m}t_{m-1}\cdots t_{i}\cdots t_{1}$ of $\left(ab\right)$
and suppose that $t_{i}=\left(x_{1}y_{1}\right)$ is an $\cutedge ab$,
as shown in Figure \ref{fig:one-cut-edge}. Let $\pi_{j}=t_{j}\cdots t_{1}$.
Since there exists a path between $a$ to $b$, there also exists
a path between $a$ and $x_{1}$ that does not use the edge $\left(x_{1}y_{1}\right)$.
Thus, in $\mathcal{M}-\left(x_{1}y_{1}\right)$, $a$ and $x_{1}$
are in the same {}``component''. Denote this component by $B_{1}$.
Similarly, a component, denoted by $B_{2}$, must contain both the
vertices $b$ and $y_{1}$. Since there is no transposition in $\pi_{i-1}$
with endpoints in both \textbf{$B_{1}$ }and $B_{2}$, there is no
element $z\in B_{1}$ such that $\pi_{i-1}\left(z\right)\in B_{2}$.
Similarly, there is no element $z\in B_{2}$ such that $\pi_{i-1}\left(z\right)\in B_{1}$.
This implies that $\pi_{i-1}\left(a\right)\in B_{1}$ and $\pi_{i-1}\left(b\right)\in B_{2}$.
Since $\left(x_{1}y_{1}\right)$ is the only edge connecting $B_{1}$
and $B_{2}$, we must have \begin{align*}
\pi_{i-1}^{-1}\left(x_{1}\right) & =a,\\
\pi_{i-1}^{-1}\left(y_{1}\right) & =b,\end{align*}
 and\begin{align*}
\pi_{i}^{-1}\left(x_{1}\right) & =b,\\
\pi_{i}^{-1}\left(y_{1}\right) & =a.\end{align*}

Now suppose there are at least two $\cutedge ab$s in $T$ as shown
in Figure \ref{fig:two-cut-edge}. Let the decomposition of $(ab)$
be $t_{m}\cdots t_{l}\cdots t_{i}\cdots t_{1}$, where $t_{l}=\left(x_{2}y_{2}\right)$
and $t_{i}=\left(x_{1}y_{1}\right)$, for some $i<l$. Define $B_{1}$,
$B_{2}$, and $B_{3}$ to be the components containing $a$, $y_{1}$,
and $y_{2}$, respectively, in $\mathcal{M}-\left(x_{1}y_{1}\right)-\left(x_{2}y_{2}\right)$.
By the same reasoning as above we must have \begin{align*}
\pi_{l-1}^{-1}\left(x_{2}\right) & =a,\\
\pi_{l-1}^{-1}\left(y_{2}\right) & =b.\end{align*}
However, this cannot be true: after applying $\left(x_{1}y_{1}\right)$,
the successor of $b$ belongs to $B_{1}$, and there are no other
edges connecting the two components of the multigraph. Hence, the
successor of $b$ before transposing $\left(x_{2}y_{2}\right)$ (that
is, the successor of $b$ in $\pi_{l-1}$) cannot be $y_{2}$.

Since $\mathcal{M}$ cannot contain more than one $\cutedge ab$,
it must contain either one $\cutedge ab$ or it must contain no $\cutedge ab$s. 

Consider next the case when there is no $\cutedge ab$ in $\mathcal{M}$.
In this case, based on Menger's theorem, there must exist at least
two pairwise edge disjoint paths between $a$ and $b$. The cost of
one of these paths has to be less than or equal to the cost of the
other path. Refer to this path as the \emph{minimum path}. Clearly,
the cost of the decomposition of $(ab)$ described by $\mathcal{M}$
is greater than or equal to twice the cost of the minimum path. 

Let the edges of the minimum path be $(az_{1})(z_{1}z_{2})...(z_{m-1}z_{m})(z_{m}b)$,
for some integer $m$. The cost of $(ab)$ is greater than or equal
to

\begin{equation}
\begin{split}
\text{\small $2\varphi^{*}(a,z_{1})+2\varphi^{*}(z_{1},z_{2})+\cdots+2\varphi^{*}(z_{m-1}z_{m})+2\varphi^{*}(z_{m},b)\ge$}\\
\text{\small $\varphi^{*}(a,z_{1})+2\varphi^{*}(z_{1},z_{2})+...+2\varphi^{*}(z_{m-1}z_{m})+2\varphi^{*}(z_{m},b)\geq$}\\
\text{\small $\varphi^{*}(a,z_{2})+2\varphi^{*}(z_{2},z_{3})+...+2\varphi^{*}(z_{m},b)\geq$}\\
\text{\small $\cdots\ge\varphi^{*}(a,z_{m})+2\varphi^{*}(z_{m},b)\geq\varphi^{*}(a,b),$}
\end{split}
\end{equation}and the cost of the decomposition associated with $\mathcal{M}$ cannot
be smaller than the cost of the optimal decomposition produced by
Alg. \ref{alg:opt-trans}. 

Next, consider the case when there is one $\cutedge ab$ in $\mathcal{M}$.
In this case, we distinguish two scenarios: when $x_{1}=a$, and when
$x_{1}\neq a$. 

In the former case, the transposition $(ay_{1})$ plays the role of
the transposition $(az_{1})$ and the remaining transpositions used
in the decomposition lie in the graph $\mathcal{M}-(ay_{1})$. Since
$\mathcal{M}-(ay_{1})$ has no $\cutedge ab$, continuing with line
two of (6) proves that the cost of the decomposition associated with
$\mathcal{M}$ cannot be smaller than $\varphi^{*}(a,b).$

In the later case, the procedure we described for the case $x_{1}=a$
is first applied to the multigraph containing the edge $(x_{1}y_{1})$
and the sub-multigraph containing the vertex $a$. As a result, the
edge $(x_{1}y_{1})$ is replaced by $(ay_{1}),$ with cost greater
than or equal to $\varphi^{*}(a,y_{1})$. Applying the same procedure
again, now for the case $x_{1}=a$, proves the claimed result.
\end{IEEEproof}
As an illustration, one can see in Figures 1a-1l that the multigraphs
corresponding to decompositions of length five have no more than one
$\cutedge ab$.

A quick inspection of Alg. 1 reveals that it has the structure of
a Viterbi-type search for finding a minimum cost path in a transposition
graph. An equivalent search procedure can be devised to operate on
the graph $\mathcal{K}\left(\varphi\right)$, rather than on a trellis.
The underlying search algorithm is described in the Appendix, and
is based on a modification of the well known Bellman-Ford procedure
\cite{cormen24introduction}.
\begin{defn}
For an arbitrary path $p=c_{1}c_{2}\cdots c_{m}$ in $\mathcal{K}\left(\varphi\right)$,
the \emph{transposition path cost }is defined as \[
\bar{\varphi}\left(p\right)=2\sum_{i=1}^{m}\varphi\left(c_{i},c_{i+1}\right)-\max_{i}\varphi\left(c_{i},c_{i+1}\right).\]
Let $\hat{p}\left(a,b\right)$ be a path with minimum transposition
path cost among paths between $a$ and $b$. That is, \[
\bar{\varphi}(\hat{p}\left(a,b\right))=\min_{p}\bar{\varphi}\left(p\right),\]
where the minimum is taken over all paths $p$ in $\mathcal{K}\left(\varphi\right)$
between $a$ and $b$. Furthermore, let $p^{*}\left(a,b\right)$ be
the standard shortest path between $a$ and $b$ in the cost graph
$\mathcal{K}\left(\varphi\right)$.\end{defn}
\begin{lem}
\label{lem:star-and-bar}The minimum cost of a transpositions $\left(ab\right)$
is at most $\bar{\varphi}\left(\hat{p}\left(a,b\right)\right)$. That
is, \[
\varphi^{*}\left(a,b\right)\le\bar{\varphi}\left(\hat{p}\left(a,b\right)\right).\]
\end{lem}
\begin{IEEEproof}
Suppose that $\hat{p}=c_{0}c_{1}\cdots c_{m}c_{m+1}$ where $a=c_{0}$
and $b=c_{m+1}$. Note that, for any $0\le i\le m$,

\begin{align}
(ab) & =\left(c_{0}c_{1}\cdots c_{i-1}c_{i}\right)\left(c_{m+1}c_{m}\cdots c_{i+2}c_{i+1}\right)\nonumber \\
 & \left(c_{i}c_{i+1}\right)\left(c_{i+1}c_{i+2}\cdots c_{m}c_{m+1}\right)\left(c_{i}c_{i-1}\cdots c_{1}c_{0}\right),\label{eq:basic-path}\end{align}
 Choose $i=\arg\max_{j}\varphi\left(c_{j},c_{j+1}\right)$ so that
$\left(c_{i}c_{i+1}\right)$ is the most costly edge in $\hat{p}\left(a,b\right)$. 

Each of the cycles in \eqref{eq:basic-path} can be decomposed using
the edges of $p$ as\begin{align}
\left(c_{0}c_{1}\cdots c_{i-1}c_{i}\right) & =\left(c_{i-1}c_{i}\right)\cdots\left(c_{2}c_{1}\right)\left(c_{1}c_{0}\right),\nonumber \\
\left(c_{i}c_{i-1}\cdots c_{1}c_{0}\right) & =\left(c_{0}c_{1}\right)\left(c_{1}c_{2}\right)\cdots\left(c_{i-1}c_{i}\right),\nonumber \\
\left(c_{i+1}c_{i+2}\cdots c_{m}c_{m+1}\right) & =\left(c_{m}c_{m+1}\right)\left(c_{m-1}c_{m}\right)\cdots\left(c_{i+2}c_{i+1}\right),\nonumber \\
\left(c_{m+1}c_{m}\cdots c_{i+2}c_{i+1}\right) & =\left(c_{i+1}c_{i+2}\right)\cdots\left(c_{m-1}c_{m}\right)\left(c_{m}c_{m+1}\right).\label{eq:basic-subs}\end{align}
Thus, the minimum cost of $\left(ab\right)$ does not exceed \[
2\sum_{j=0}^{m}\varphi\left(c_{j},c_{j+1}\right)-\varphi\left(c_{i},c_{i+1}\right).\]
\end{IEEEproof}
\begin{lem}
\label{lem:Transpositional-path}The minimum cost of a transposition
$\left(ab\right)$ equals the minimum transposition path cost $\varphi(\hat{p}\left(a,b\right))$.
That is,\[
\varphi^{*}\left(a,b\right)=\bar{\varphi}\left(\hat{p}\left(a,b\right)\right).\]
\end{lem}
\begin{IEEEproof}
Suppose $\tau$ is the minimum cost decomposition of $\left(ab\right)$.
Let $\mathcal{M}=\mathcal{T}\left(\tau\right)$, and note that $\cost\left(\mathcal{M}\right)=\varphi^{*}\left(a,b\right)$. 

In the proof of Theorem \ref{thm:correctness_of_alg1}, we showed
that $\mathcal{M}$ has at most one $\cutedge ab$. Suppose that $\mathcal{M}$
has no $\cutedge ab$. Then there are two edge-disjoint paths between
$a$ and $b$ in $\mathcal{M}$. Define the minimum path as in Theorem
\ref{thm:correctness_of_alg1}. Suppose the minimum path is $p=c_{0}c_{1}\cdots c_{m}c_{m+1}$.
It is easy to see that \[
\bar{\varphi}\left(\hat{p}\left(a,b\right)\right)\le\bar{\varphi}\left(p\right)\le\cost\left(\mathcal{M}\right)=\varphi^{*}\left(a,b\right).\]
From Lemma \eqref{lem:star-and-bar}, we have $\varphi^{*}\left(a,b\right)\le\bar{\varphi}\left(\hat{p}\left(a,b\right)\right)$.
Hence, $ $in this case, we conclude that $\varphi^{*}\left(a,b\right)=\bar{\varphi}\left(\hat{p}\left(a,b\right)\right)$.

Next, suppose that $\mathcal{M}$ has one $\cutedge ab$, as shown
in Figure \ref{fig:one-cut-edge}. Menger's theorem implies that there
are two edge-disjoint paths between $a$ and $x_{1}$ and two edge-disjoint
paths between $b$ and $y_{1}$. Let $p_{1}$ be the path with smaller
cost among the pair of paths between $a$ and $x_{1}$, and similarly,
let $p_{2}$ be the path with smaller cost between the pair of paths
between $b$ and $y_{1}$. Let $p$ be the path obtained by concatenating
$p_{1}$, the edge $\left(x_{1}y_{1}\right)$, and $p_{2}$. Note
that $\bar{\varphi}\left(\hat{p}\left(a,b\right)\right)\le\bar{\varphi}\left(p\right)\le\cost\left(\mathcal{M}\right)\le\varphi^{*}\left(a,b\right)$.
Since $\varphi^{*}\left(a,b\right)\le\bar{\varphi}\left(\hat{p}\left(a,b\right)\right)$,
we have $\varphi^{*}\left(a,b\right)=\bar{\varphi}\left(\hat{p}\left(a,b\right)\right)$.
\end{IEEEproof}
It is easy to see that $\varphi^{*}\left(a,b\right)\le2\cost\left(p^{*}\left(a,b\right)\right)$
since we have \begin{align}
\varphi^{*}\left(a,b\right) & =\bar{\varphi}\left(\hat{p}\left(a,b\right)\right)\nonumber \\
 & \le\bar{\varphi}\left(p^{*}\left(a,b\right)\right)\nonumber \\
 & \le2\cost\left(p^{*}\left(a,b\right)\right).\label{eq:p-hat-p}\end{align}

Note that the Bellman-Ford Alg. \ref{alg:SSBF}, presented in the
Appendix, finds the paths $\hat{p}$ in $\mathcal{K}(\varphi)$ between
a given vertex $s$ and all other vertices in the graph.

Lemma \ref{lem:Transpositional-path} provides an easy method for
computing $\varphi^{*}\left(i,j\right)$ when there is only one path
with finite cost between $i$ and $j$ in $\mathcal{K}\left(\varphi\right)$.
For example, for an extended-metric cost function $\varphi$, we have
\begin{equation}
\varphi^{*}\left(i,j\right)=2\sum_{t=1}^{l}\varphi\left(c_{t},c_{t+1}\right)-\max_{1\le t\le l}\varphi\left(c_{t},c_{t+1}\right),\label{eq:def-navin}\end{equation}
where $\hat{p}\left(i,j\right)=c_{1}\cdots c_{l+1}$, $c_{1}=i$,
$c_{l+1}=j$, is the unique path between $i$ and $j$ in $\Theta_{s}$.

\section{Optimizing Individual Cycles\label{sec:Minimum-Cost-MLDs}}

\begin{center}
\begin{figure*}
\subfloat[$\pi_{a}=\imath=1234$]{\centering
\psset{unit=30pt}
\begin{pspicture}[showgrid=false](-1.5,-1.5)(1.5,1.5)
\cnode*(0.000000,1.000000){2pt}{N1}

\nput{90}{N1}{1}

\nccircle[nodesep=3pt]{->}{N1}{.4}

\cnode*(1.000000,0.000000){2pt}{N2}

\nput{0}{N2}{2}

\nccircle[nodesep=3pt]{->}{N2}{.4}

\cnode*(0.000000,-1.000000){2pt}{N3}

\nput{-90}{N3}{3}

\nccircle[nodesep=3pt]{->}{N3}{.4}

\cnode*(-1.000000,-0.000000){2pt}{N4}

\nput{-180}{N4}{4}

\nccircle[nodesep=3pt]{->}{N4}{.4}

\end{pspicture}

\label{fig:1a}}\subfloat[$\pi_{b}=2134$]{\centering
\psset{unit=30pt}
\begin{pspicture}[showgrid=false](-1.75,-1.75)(1.75,1.75)
\cnode*(0.000000,1.000000){2pt}{N1}

\nput{90}{N1}{1}

\cnode*(1.000000,0.000000){2pt}{N2}

\nput{0}{N2}{2}

\cnode*(0.000000,-1.000000){2pt}{N3}

\nput{-90}{N3}{3}

\cnode*(-1.000000,-0.000000){2pt}{N4}

\nput{-180}{N4}{4}

\ncarc[arcangle=-45]{->}{N1}{N2}

\ncarc[arcangle=-45]{->}{N2}{N1}

\ncarc[arcangle=-45]{->}{N3}{N4}

\ncarc[arcangle=-45]{->}{N4}{N3}

\ncarc[arcangle=0,linestyle=dashed,linewidth=.2pt]{->}{N1}{N2}

\end{pspicture}

\label{fig:1b}}\subfloat[$\pi_{c}=3142$]{\centering
\psset{unit=30pt}
\begin{pspicture}[showgrid=false](-1.75,-1.75)(1.75,1.75)
\cnode*(0.000000,1.000000){2pt}{N1}

\nput{90}{N1}{1}

\cnode*(1.000000,0.000000){2pt}{N2}

\nput{0}{N2}{2}

\cnode*(0.000000,-1.000000){2pt}{N3}

\nput{-90}{N3}{3}

\cnode*(-1.000000,-0.000000){2pt}{N4}

\nput{-180}{N4}{4}

\ncarc[arcangle=0]{->}{N4}{N2}

\ncarc[arcangle=-45]{->}{N2}{N1}

\ncarc[arcangle=45]{->}{N3}{N4}

\ncarc[arcangle=0]{->}{N1}{N3}

\ncarc[arcangle=0,linestyle=dashed,linewidth=.2pt]{->}{N1}{N2}

\ncarc[arcangle=0,linestyle=dashed,linewidth=.2pt]{->}{N2}{N3}

\end{pspicture}

\label{fig:1c}}\subfloat[$\pi_{d}=3241$]{\centering
\psset{unit=30pt}
\begin{pspicture}[showgrid=false](-1.75,-1.75)(1.75,1.75)
\cnode*(0.000000,1.000000){2pt}{N1}

\nput{90}{N1}{1}

\cnode*(1.000000,0.000000){2pt}{N2}

\nput{0}{N2}{2}

\cnode*(0.000000,-1.000000){2pt}{N3}

\nput{-90}{N3}{3}

\cnode*(-1.000000,-0.000000){2pt}{N4}

\nput{-180}{N4}{4}

\nccircle[nodesep=3pt]{->}{N2}{.4}

\ncarc[arcangle=45]{->}{N4}{N1}

\ncarc[arcangle=45]{->}{N3}{N4}

\ncarc[arcangle=0]{->}{N1}{N3}

\ncarc[arcangle=0,linestyle=dashed,linewidth=.2pt]{->}{N1}{N2}

\ncarc[arcangle=0,linestyle=dashed,linewidth=.2pt]{->}{N2}{N3}

\end{pspicture}

\label{fig:1d}}\subfloat[$\pi_{e}=4231$]{\centering
\psset{unit=30pt}
\begin{pspicture}[showgrid=false](-1.75,-1.75)(1.75,1.75)
\cnode*(0.000000,1.000000){2pt}{N1}

\nput{90}{N1}{1}

\cnode*(1.000000,0.000000){2pt}{N2}

\nput{0}{N2}{2}

\cnode*(0.000000,-1.000000){2pt}{N3}

\nput{-90}{N3}{3}

\cnode*(-1.000000,-0.000000){2pt}{N4}

\nput{-180}{N4}{4}

\nccircle[nodesep=3pt]{->}{N2}{.4 }

\ncarc[arcangle=45]{->}{N4}{N1}

\ncarc[arcangle=45]{->}{N1}{N4}

\nccircle[nodesep=3pt]{->}{N3}{.4}

\ncarc[arcangle=0,linestyle=dashed,linewidth=.2pt]{->}{N1}{N2}

\ncarc[arcangle=0,linestyle=dashed,linewidth=.2pt]{->}{N2}{N3}

\ncarc[arcangle=0,linestyle=dashed,linewidth=.2pt]{->}{N3}{N4}

\end{pspicture}

\label{fig:1e}}

\caption{In any decomposition of the transposition $\left(14\right)$, there
exists a path from $1$ to $4$. In this illustration, the iteratively
added edges of the path are dashed.}
\label{fig:path-a-b-(ab)}
\end{figure*}
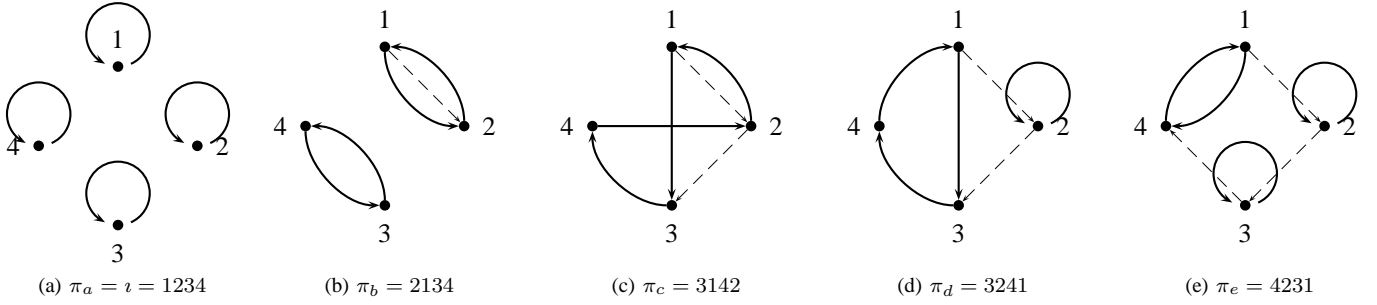

\par\end{center}

We consider next the cost optimization problem over single cycles.
First, we find the minimum cost MLD via a dynamic programming algorithm.
The minimum cost MLD is obtained with respect to the optimized cost
function $\varphi^{*}$ of the previous section. For simplicity, we
henceforth omit the superscript in the cost whenever there is no ambiguity
in terms which cost function is used. 

We also present a second algorithm to find decompositions whose cost,
along with the cost of the minimum cost MLD, is not more than a constant
factor higher than the cost of the MCD. Both algorithms are presented
for completeness.

The results in this section apply to any cycle $\sigma$. However,
for clarity of presentation, and without loss of generality, we consider
the cycle $\sigma=\left(12\cdots k\right)$.

\subsection{Minimum Cost, Minimum Length Transposition Decomposition}

Recall that the vertices of $\mathcal{G}\left(\sigma\right)$ are
placed on a circle. For an MLD $\tau$ of a permutation $\pi$ with
$\ell$ cycles, $\mathcal{T}\left(\tau\right)$ is a forest with $\ell$
components; each tree in the forest is the decomposition of one cycle
of $\pi$. This can be easily seen by observing that each cycle corresponds
to a tree. The following lemma provides a rigorous proof for this
statement.
\begin{lem}
The graph $\mathcal{T}\left(\tau\right)$ of an MLD $\tau$ of a cycle
$\sigma$ is a tree.\label{lem:MLD-graph}\end{lem}
\begin{IEEEproof}
First, we show that $\mathcal{T}\left(\tau\right)$ is connected.
The decomposition $\tau$ transform the identity permutation $\imath$
to $\sigma$ by transposing pairs of elements. Note that every transposition
exchanges the predecessors of two elements. In $\imath$, each element
$i$ is a fixed point (i.e., it is its own predecessor) and in $\sigma$,
$i$ is the predecessor of $\sigma\left(i\right)$. Thus there exists
a path, formed by a sequence of transpositions, between $i$ and $\sigma\left(i\right)$.
An instance of such a path is described in Example \ref{exa:path-from-a-to-b-(ab)}.

To complete the proof, observe that $\mathcal{T}\left(\tau\right)$
has $k$ vertices and $k-1$ edges, since an MLD of a cycle of length
$k$ contains $k-1$ transpositions. Hence, $\mathcal{T}\left(\tau\right)$
is a tree.
\end{IEEEproof}
As already pointed out, we provide an example that illustrates the
existence of a path from $i$ to $\sigma\left(i\right)$ in the decomposition
$\tau$ of $\sigma$, for the special case when $\sigma$ is a cycle
of length two.
\begin{example}
\label{exa:path-from-a-to-b-(ab)}Consider the cycle $\sigma=\left(14\right)$.
It is easy to see that $\tau=\left(34\right)\left(12\right)\left(23\right)\left(34\right)\left(12\right)$
is a decomposition of $\sigma$. Figure \ref{fig:path-a-b-(ab)} illustrates
a path from vertex 1 to vertex 4 in $\mathcal{T}\left(\tau\right)$.
For instance, the transposition $\left(12\right)$ in $\tau$ corresponds
to the edge $\left(12\right)$ in $\mathcal{T}\left(\tau\right)$,
as shown in Figure \ref{fig:1b}, and the transposition $\left(23\right)$
corresponds to the edge $\left(23\right)$ etc. The cycle $\sigma$
is a cycle in $\pi_{e}$ in Figure \ref{fig:1e}. The path from $1$
to $4$ in $\mathcal{T}\left(\tau\right)$ is $1\rightarrow2\rightarrow3\rightarrow4$.\queede
\end{example}
For related ideas regarding permutation decompositions and graphical
structures, the interested reader is referred to \cite{goulden_tree-like_2002}.

The following definitions will be used in the proof of a lemma which
states that $\mathcal{G}\left(\pi\right)\cup\mathcal{T}\left(\tau\right)$
is planar, provided that $\tau$ is an MLD of $\sigma$.

Let $R$ be the region enclosed by edges of $\mathcal{G}\left(\sigma\right)$.
Let $T$ be a tree with vertex set $\{1,2,\cdots,k\}$, such that
$\mathcal{G}\left(\sigma\right)\cup T$ is planar. Since $T$ is a
tree with edges contained in $R$, the edges of $T$ partition $R$
into smaller regions; each of these parts is the enclosure of a subset
of edges of $\mathcal{G}\left(\sigma\right)\cup T$ and includes the
vertices of these edges. These vertices can be divided into \emph{corner
vertices,} lying at the intersection of at least two regions, and
\emph{inner vertices}, belonging only to one region. In Figure \ref{fig:regions},
$\mathcal{G}\left(\sigma\right)$ with vertices $V=\left\{ 1,2,3,4,5,6\right\} $
is partitioned into four  regions, $R_{1},R_{2},R_{3}$ and $R_{4}$.
In $R_{2}$, vertices $1$ and $3$ are corner vertices, while vertex
$2$ is an inner vertex.
\begin{lem}
\label{lem:MLD-planar} For an MLD $\tau=t_{1}\cdots t_{k-1}$ of
$\sigma$, $\mathcal{T}\left(\tau\right)\cup\mathcal{G}\left(\sigma\right)$
is planar. That is, for $t_{i}=\left(a_{1}a_{2}\right)$, where $a_{1}<a_{2}$,
and $t_{j}=\left(b_{1}b_{2}\right)$, where $b_{1}<b_{2}$, if $a_{1}<b_{1}<a_{2}$,
then $a_{1}<b_{2}<a_{2}$. \end{lem}
\begin{IEEEproof}
Note that $\tau^{-1}\sigma=\imath$. Let $\tau_{i}=t_{i-1}\cdots t_{1}$.
Since $\tau$ is an MLD of $\sigma$, $\tau_{i}\sigma$ has $i$ cycles.
The proof proceeds by showing that for all $1\le i\le k$, the following
two claims are true: 

(I) $\mathcal{G}\left(\sigma\right)\cup\mathcal{T}\left(\tau_{i}\right)$
is planar. 

(II) Each cycle of $\tau_{i}$ corresponds to a subregion $R$ of
$\mathcal{G}\left(\sigma\right)\cup\mathcal{T}\left(\tau_{i}\right)$.
The cycle corresponding to $R$ contains all of its inner vertices
and some of its corner vertices but no other vertex.

Both claims (I) and (II) are obvious for $i=1$. We show that if (I)
and (II) are true for $\tau_{i}$, then they are also true for $\tau_{i+1}$.

Let $t_{i}=\left(ab\right)$. Clearly, $\tau_{i+1}\sigma=t_{i}\tau_{i}\sigma$
has one more cycle than $\tau_{i}\sigma$, and by assumption, $\mathcal{G}\left(\sigma\right)\cup\mathcal{T}\left(\tau_{i}\right)$
is planar and partitioned into a set of subregions. Note that $a$
and $b$ are in the same cycle, and thus are inner or corner veritices
of some subregion $R^{*}$ of $\mathcal{G}\left(\sigma\right)\cup\mathcal{T}\left(\tau_{i}\right)$.
The edge $\left(ab\right)$ divides $R^{*}$ into two subregions,
$R_{a}$ and $R_{b}$ (without crossing any edge in $\mathcal{G}\left(\sigma\right)\cup\mathcal{T}\left(\tau_{i}\right)$).
This proves (I).

Let the cycle corresponding to $R^{*}$ be \[
\mu=\left(aa_{1}\cdots a_{l}bb_{1}\cdots b_{l'}\right)\]
as seen in Figure \ref{fig:ab-cycle-b}. Then, \[
\left(ab\right)\mu=\left(aa_{1}\cdots a_{l}\right)\left(bb_{1}\cdots b_{l'}\right).\]
Now the cycles $\left(aa_{1}\cdots a_{l}\right)$ and $\left(bb_{1}\cdots b_{l'}\right)$
in $\tau_{i+1}$ correspond to subregions $R_{a}$ and $R_{b}$, respectively,
as see in Figure \ref{fig:ab-cycle-c}. This proves claim (II) since
the cycle corresponding to each subregion contains all of its inner
vertices and some of its corner vertices but no other vertex.
\end{IEEEproof}
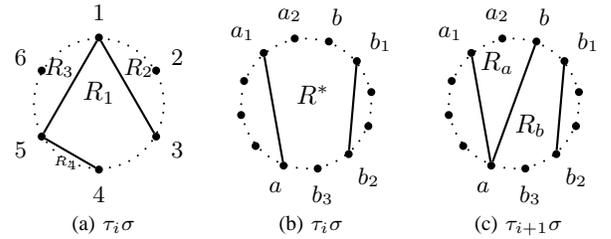
\begin{figure}
\begin{centering}
\subfloat[$\tau_{i}\sigma$]{\centering
\psset{unit=25pt}
\begin{pspicture}[showgrid=false](-1.5,-1.5)(1.5,1.5)
\cnode*(0.000,1){1.5pt}{N1}

\nput{90}{N1}{\small$1$}

\cnode*(0.866,0.5){1.5pt}{N2}

\nput{30}{N2}{\small$2$}

\cnode*(0.866,-0.5){1.5pt}{N3}

\nput{-30}{N3}{\small$3$}

\cnode*(0.000,-1){1.5pt}{N4}

\nput{-90}{N4}{\small$4$}

\cnode*(-0.866,-0.5){1.5pt}{N5}

\nput{-150}{N5}{\small$5$}

\cnode*(-0.866,0.5){1.5pt}{N6}

\nput{-210}{N6}{\small$6$}

\pscircle[linestyle=dotted](0,0){1}

\ncline{-}{N1}{N3}

\ncline{-}{N4}{N5}

\ncline{-}{N1}{N5}

\rput[b]{0}(0,0){$R_1$}

\rput[b]{0}(.6,0.4){\small $R_2$}

\rput[b]{0}(-.6,.4){\small $R_3$}

\rput[b]{0}(-.5,-1){\tiny $R_4$}

\end{pspicture}

\label{fig:regions}}\subfloat[$\tau_{i}\sigma$]{\centering
\psset{unit=25pt}
\begin{pspicture}[showgrid=false](-1.5,-1.5)(1.5,1.5)
\cnode*(0.985,0.17364817766693){1.5pt}{N1}

\nput{10}{N1}{\small$$}

\cnode*(0.766,0.642787609686539){1.5pt}{N2}

\nput{40}{N2}{\small$b_1$}

\cnode*(0.342,0.939692620785908){1.5pt}{N3}

\nput{70}{N3}{\small$b$}

\cnode*(-0.174,0.984807753012208){1.5pt}{N4}

\nput{100}{N4}{\small$a_2$}

\cnode*(-0.643,0.766044443118978){1.5pt}{N5}

\nput{130}{N5}{\small$a_1$}

\cnode*(-0.940,0.342020143325669){1.5pt}{N6}

\nput{160}{N6}{\small$$}

\cnode*(-0.985,-0.17364817766693){1.5pt}{N7}

\nput{190}{N7}{\small$$}

\cnode*(-0.766,-0.642787609686539){1.5pt}{N8}

\nput{220}{N8}{\small$$}

\cnode*(-0.342,-0.939692620785908){1.5pt}{N9}

\nput{250}{N9}{\small$a$}

\cnode*(0.174,-0.984807753012208){1.5pt}{N10}

\nput{280}{N10}{\small$b_3$}

\cnode*(0.643,-0.766044443118978){1.5pt}{N11}

\nput{310}{N11}{\small$b_2$}

\cnode*(0.940,-0.342020143325669){1.5pt}{N12}

\nput{340}{N12}{\small$$}

\pscircle[linestyle=dotted](0,0){1}

\ncline{-}{N2}{N11}

\ncline{-}{N5}{N9}

\rput[b]{0}(0.1,0){$R^*$}

\end{pspicture}

\label{fig:ab-cycle-b}}\subfloat[$\tau_{i+1}\sigma$]{\centering
\psset{unit=25pt}
\begin{pspicture}[showgrid=false](-1.5,-1.5)(1.5,1.5)
\cnode*(0.985,0.17364817766693){1.5pt}{N1}

\nput{10}{N1}{\small$$}

\cnode*(0.766,0.642787609686539){1.5pt}{N2}

\nput{40}{N2}{\small$b_1$}

\cnode*(0.342,0.939692620785908){1.5pt}{N3}

\nput{70}{N3}{\small$b$}

\cnode*(-0.174,0.984807753012208){1.5pt}{N4}

\nput{100}{N4}{\small$a_2$}

\cnode*(-0.643,0.766044443118978){1.5pt}{N5}

\nput{130}{N5}{\small$a_1$}

\cnode*(-0.940,0.342020143325669){1.5pt}{N6}

\nput{160}{N6}{\small$$}

\cnode*(-0.985,-0.17364817766693){1.5pt}{N7}

\nput{190}{N7}{\small$$}

\cnode*(-0.766,-0.642787609686539){1.5pt}{N8}

\nput{220}{N8}{\small$$}

\cnode*(-0.342,-0.939692620785908){1.5pt}{N9}

\nput{250}{N9}{\small$a$}

\cnode*(0.174,-0.984807753012208){1.5pt}{N10}

\nput{280}{N10}{\small$b_3$}

\cnode*(0.643,-0.766044443118978){1.5pt}{N11}

\nput{310}{N11}{\small$b_2$}

\cnode*(0.940,-0.342020143325669){1.5pt}{N12}

\nput{340}{N12}{\small$$}

\pscircle[linestyle=dotted](0,0){1}

\ncline{-}{N2}{N11}

\ncline{-}{N5}{N9}

\ncline{-}{N3}{N9}

\rput[b]{0}(-.25,.45){$R_a$}

\rput[b]{0}(0.25,-.5){$R_b$}

\end{pspicture}

\label{fig:ab-cycle-c}}
\par\end{centering}

\caption{A region is divided into two regions by transposition $\left(ab\right)$.
See proof of Lemma \ref{lem:MLD-planar}.}

\label{fig:ab-cycle}
\end{figure}

The following lemma establishes a partial converse to the previous
lemma.
\begin{lem}
\label{lem:Planar-MLD} For a cycle $\sigma$ and a spanning tree
$T$ over the vertices $\left\{ 1,2,\cdots,k\right\} $, for which
$\mathcal{G}\left(\sigma\right)\cup T$ is planar, there exists at
least one MLD $\tau$ of $\sigma$ such that $T=\mathcal{T}\left(\tau\right)$.\end{lem}
\begin{IEEEproof}
We prove the lemma by recursively constructing an MLD corresponding
to $T$. If $k=2$, then $T$ has exactly one edge and the MLD is
the transposition corresponding to that edge. For $k>2$, some vertex
has degree larger than one. Without loss of generality, assume that
$\deg\left(1\right)>1$. Let\[
r=\max\left\{ u|\left(1u\right)\in T\right\} .\]
Since $T$ is a tree, $T-\left(1r\right)$ has two components. These
two components have vertex sets $\left\{ 1,\cdots,s\right\} $ and
$\left\{ s+1,\cdots,k\right\} $, for some $s$. It is easy to see
that \begin{equation}
\left(1\,\cdots\, k\right)=\left(s+1\,\cdots\, k\,1\right)\left(1\,\cdots\, s\right).\label{eq:two-cycle}\end{equation}
 Let\begin{eqnarray*}
T' & = & T\left[\left\{ 1,\cdots,s\right\} \right],\\
T'' & = & T\left[\left\{ s+1,\cdots,k,1\right\} \right].\end{eqnarray*}
Note that $T'$ and $T''$ have fewer than $k$ vertices. Furthermore,
$T'\cup\mathcal{G}\left(\left(1\cdots\, s\right)\right)$ and $T''\cup\mathcal{G}\left(\left(s+1\cdots k\,1\right)\right)$
are planar. Thus, from the induction hypothesis, $\left(1\,\cdots\, s\right)$
and $\left(s+1\,\cdots\, k\,1\right)$ have decompositions $\tau'$
and $\tau''$ of length $s-1$ and $k-s$, respectively. By  \eqref{eq:two-cycle},
$\tau''\tau'$ is an MLD for $\sigma$.\end{IEEEproof}
\begin{example}
In Figure~\ref{fig:no-last-edge-1}, we have $r=10$ and $s=8$.
The cycle $\left(1\,\cdots\,12\right)$ can be decomposed  into two
cycles,\[
\left(1\,\cdots\,12\right)=\left(9\,10\,11\,12\,1\right)\left(1\,2\,\cdots\,8\right).\]
Now, each of these cycles is decomposed into smaller cycles, for example
\begin{align*}
\left(9\,10\,11\,12\,1\right) & =\left(9\,10\right)\left(10\,11\,12\,1\right),\\
\left(1\cdots8\right) & =\left(8\,1\right)\left(1\cdots7\right).\end{align*}
The same type of decomposition can be performed on the cycles $\left(9\,10\right),\cdots,\left(17\right)$.
\queede
\end{example}
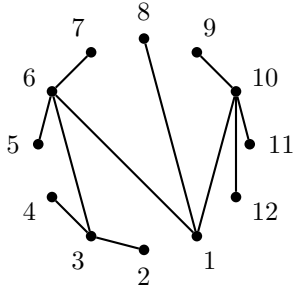
\begin{figure}
\begin{center}
\psset{unit=40pt}
\begin{pspicture}[showgrid=false](-2,-2)(2,2) 
\cnode* (0.00,1.00) {2pt} {n8}

\nput  {90}{n8}{$8$} 

\cnode* (0.50,0.87) {2pt}	{n9}

\nput  {60}{n9}{$9$} 

\cnode* (0.87,0.50) {2pt}	{n10}

\nput  {30}{n10}{$10$} 

\cnode* (1.00,0.00) {2pt}	{n11}

\nput  {0}{n11}{$11$} 

\cnode* (0.87,-0.50) {2pt}	{n12}

\nput  {330}{n12}{$12$} 

\cnode* (0.50,-0.87) {2pt}	{n1}

\nput  {300}{n1}{$1$} 

\cnode* (0.00,-1.00) {2pt}	{n2}

\nput  {270}{n2}{$2$} 

\cnode*(-0.50,-0.87) {2pt}	 {n3}

\nput  {240}{n3}{$3$} 

\cnode* (-0.87,-0.50) {2pt}	{n4}

\nput  {210}{n4}{$4$} 

\cnode*(-1.00,-0.00) {2pt}	 {n5}

\nput  {180}{n5}{$5$} 

\cnode*(-0.87,0.50) {2pt}	 {n6}

\nput  {150}{n6}{$6$} 

\cnode*(-0.50,0.87) {2pt}	 {n7}

\nput  {120}{n7}{$7$} 

\ncline{-}{n8}{n1} \ncline{-}{n9}{n10} 

\ncline{-}{n10}{n12} \ncline{-}{n11}{n10} 

\ncline{-}{n10}{n1} \ncline{-}{n6}{n7} 

\ncline{-}{n6}{n1} \ncline{-}{n2}{n3} 

\ncline{-}{n3}{n4} \ncline{-}{n3}{n6} 

\ncline{-}{n5}{n6}
\end{pspicture}
\end{center}\caption{Example illustrating the proof of Lemma \ref{lem:Planar-MLD}: $r=10,s=8$.}
\label{fig:no-last-edge-1}
\end{figure}

Since any MLD of a cycle can be represented by a tree that is planar
on the circle, the search for an MLD of minimum cost only needs to
be performed over the set of planar trees. This search can be performed
using a dynamic program, outlined in Alg. \ref{alg:opt-cycle}. Lemma
\ref{lem:opt-cycle} establishes that Alg. \ref{alg:opt-cycle} produces
a minimum cost MLD.

\begin{algorithm}
\caption{\textsc{Min-Cost-MLD}}
\label{alg:opt-cycle}\begin{algorithmic}[1]
\State Input: Optimized transposition cost function $\Phi^*$ where $\Phi^*_{i,j}=\varphi^*(i,j)$ (Output of Alg. \ref{alg:opt-trans})
\State $C(i,j)\leftarrow\infty $ for $i,j\in[k]$
\State $C\left(i,i\right)\leftarrow 0$ for $i\in \left[k\right]$
\State $C\left(i,i+1\right)\leftarrow \varphi^*\left(i,i+1\right)$ for $i\in \left[k\right]$
\For {$l =2\cdots k-1$}
	\For {$i=1\cdots k-l $} 
		\State $j \leftarrow i+l $ 	
		\For {$i\le s < r \le j$}
            \State $A\leftarrow	C(i,s)+C(s+1,r)+C(r,j )+\varphi^*(i,r) $
			\If { $A < C(i,j)$}		
				\State $C(i,j)\leftarrow A$
			\EndIf
		\EndFor 	
	\EndFor 
\EndFor
\end{algorithmic}

\end{algorithm}

\begin{lem}
\label{lem:opt-cycle}The output cost of Alg. \eqref{alg:opt-cycle},
$C\left(1,k\right)$, equals $L\left(\sigma\right)$.\end{lem}
\begin{IEEEproof}
The algorithm finds the minimum cost MLD of $\left(1\cdots k\right)$
by first finding the minimum cost of MLDs of shorter cycles of the
form $\left(i\cdots j\right)$, where $1\le i<j\le k$. We look at
the computations performed in the algorithm from a top-down point
of view.

Let $C_{T}\left(i,j\right)$ be the cost of the decomposition of the
cycle $\sigma^{i,j}=\left(i\cdots j\right)$, using edges of $T\left[\left\{ i,\cdots,j\right\} \right]$,
where $T$ is an arbitrary planar spanning tree over the vertices
$\left\{ 1,\cdots,k\right\} $ arranged on a circle. For a fixed $T$,
let $r$ and $s$ be defined as in the proof of Lemma \ref{lem:Planar-MLD}.
We may write\begin{equation}
\left(i\cdots j\right)=\left(s+1\cdots r\right)\left(ir\right)\left(r\cdots j\right)\left(i\cdots s\right)\label{eq:CT0}\end{equation}
where $i\le s<r\le j$. Thus \begin{align}
C_{T}\left(i,j\right) & =C_{T}\left(s+1,r\right)+\varphi^{*}\left(i,r\right)+C_{T}\left(r,j\right)+C_{T}\left(i,s\right).\label{eq:CT-1}\end{align}
Define $C\left(i,j\right)=C_{T^{*}}\left(i,j\right)$, where \[
T^{*}=\arg\min_{T}C_{T}\left(i,j\right)\]
denotes a tree that minimizes the cost of the decomposition of $\left(i\cdots j\right)$.
Then, we have\begin{equation}
C\left(i,j\right)=C\left(s^{*}+1,r^{*}\right)+\varphi^{*}\left(i,r^{*}\right)+C\left(r^{*},j\right)+C\left(i,s^{*}\right),\label{eq:CT1}\end{equation}
where $s^{*}$ and $r^{*}$ are the values that minimize the right-hand-side
of \eqref{eq:CT-1} under the constraint $1\le i\le s<r\le j$. Since
the cost of each cycle can be computed from the cost of shorter cycles,
$C\left(i,j\right)$ can be obtained recursively, with initialization
\begin{equation}
C\left(i,i+1\right)=\varphi^{*}\left(i,i+1\right).\label{eq:CT2}\end{equation}
The algorithm searches over $s$ and $r$ and computes $C\left(1,k\right)$
using \eqref{eq:CT1} and \eqref{eq:CT2}. 

Although these formulas are written in a recursive form, Alg. \ref{alg:opt-cycle}
is written as a dynamic program. The algorithm first computes $C\left(i,j\right)$
for small values of $i$ and $j$, and then finds the cost of longer
cycles. That is, for each $2\le l\le k-1$ in increasing order, $C\left(i,i+l\right)$
is computed by choosing its optimal decomposition in terms of costs
of smaller cycles.\end{IEEEproof}
\begin{example}
As an example, let us find the minimum cost decomposition of the cycle
$\sigma=(1234)$ using the above algorithm. Let $\Phi$ be the matrix
of transposition costs, with $\Phi_{ij}=\varphi\left(i,j\right)$:
\begin{equation}
\Phi=\left[\begin{array}{cccc}
0 & 5 & 10 & 3\\
- & 0 & 2 & 3\\
- & - & 0 & 9\\
- & - & - & 0\end{array}\right],\label{eq:cost_raw}\end{equation}
\[
\Phi^{*}=\left[\begin{array}{cccc}
0 & 5 & 9 & 3\\
- & 0 & 2 & 3\\
- & - & 0 & 7\\
- & - & - & 0\end{array}\right]\]
 After optimizing the transposition costs in $\Phi$ via Alg. \ref{alg:opt-trans},
we obtain $\Phi^{*}$, shown beneath $\Phi$. From Alg. \ref{alg:opt-cycle},
we obtain\begin{align*}
C\left(1,3\right) & =C\left(2,3\right)+\varphi^{*}\left(1,2\right)=7,\left(s,r\right)=\left(1,2\right),\\
C\left(2,4\right) & =C\left(2,3\right)+\varphi^{*}\left(2,4\right)=5,\left(s,r\right)=\left(3,4\right).\end{align*}
Consider the cycle $\left(1234\right)$, where $i=1$ and $j=4$.
The algorithm compares $\binom{4}{2}=6$ ways to represent the cost
of this cycle using the cost of shorter cycles. The minimum cost is
obtained by choosing $s=2$ and $r=4$, so that\[
C\left(1,4\right)=C\left(2,4\right)+\varphi^{*}\left(1,4\right)=8.\]
Writing $C$ as a matrix, where $C\left(i,j\right)=C_{ij}$, we have:
\[
C=\left[\begin{array}{cccc}
0 & 5 & 7 & 8\\
- & 0 & 2 & 5\\
- & - & 0 & 7\\
- & - & - & 0\end{array}\right]\]
\queede
\end{example}
Note that we can modify the above algorithm to also find the underlying
MLD by using \eqref{eq:CT0} to write the decomposition of every cycle
with respect to $r$ and $s$ that minimize the cost of the cycle.
For example, from \eqref{eq:CT0}, by substituting the appropriate
values of $r$ and $s$, we obtain \begin{align*}
\left(1234\right) & =\left(234\right)\left(14\right)\\
 & =\left(34\right)\left(24\right)\left(14\right).\end{align*}

The initialization steps are performed in $O\left(k\right)$ time.
The algorithm performs a constant number of steps for each $i,j,r$,
and $s$ such that $1\le i\le s<r\le j\le k$. Hence, the computational
cost of the algorithm is $O\left(k^{4}\right)$.

\begin{figure}
\includegraphics[width=3.4in]{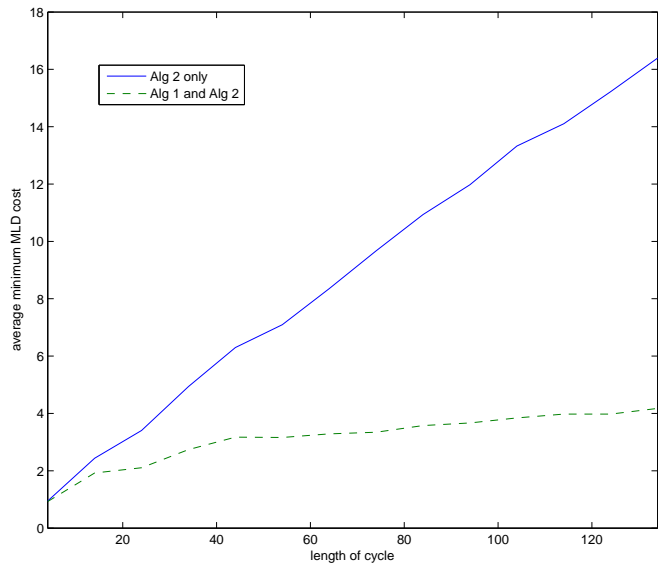} \caption{The average minimum MLD cost vs the length of the cycle. Transposition
costs are chosen independently and uniformly in {[}0,1{]}.}

\label{fig:f1} 
\end{figure}

Note that Alg. \ref{alg:opt-cycle} operates on the optimized cost
function $\varphi^{*}$, obtained as the output of Alg. \ref{alg:opt-trans}.
Figure \ref{fig:f1} illustrates the importance of first reducing
individual transposition costs using Alg. \ref{alg:opt-trans} before
applying the dynamic program. Since the dynamic program can only use
$k-1$ transpositions of minimum cost, it cannot optimize the individual
costs of transpositions and strongly relies on the reduction of Alg.
\ref{alg:opt-trans} for producing low cost solutions. In Figure \ref{fig:f1},
the transposition costs were chosen independently from a uniform distribution
over $\left[0,1\right]$. 

\subsection{Constant-factor approximation for cost of MCD}

For the cycle $\sigma=\left(12\cdots k\right)$ and $1\le j\le k$,
consider the decomposition \begin{multline*}
\left(j+1\ j+2\right)\left(j+2\ j+3\right)\cdots\\
\left(k-1\ k\right)\left(k1\right)\left(12\right)\left(23\right)\cdots\left(j-1\ j\right).\end{multline*}
 The cost of this decomposition equals \[
\sum_{i\in\sigma}\varphi^{*}\left(i,\sigma\left(i\right)\right)-\varphi^{*}\left(j,\sigma\left(j\right)\right).\]
 To minimize the cost of the decomposition, we choose $j$ such that
the transpositions $\left(j\ j+1\right)$ has maximum cost. This choice
leads to the decomposition\begin{multline}
\left(j^{*}+1\ j^{*}+2\right)\left(j^{*}+2\ j^{*}+3\right)\cdots\\
\left(k-1\ k\right)\left(k1\right)\left(12\right)\left(23\right)\cdots\left(j^{*}-1\ j^{*}\right)\label{eq:j*}\end{multline}
where \[
j^{*}=\arg\max_{j\in\sigma}\varphi^{*}\left(j,\sigma\left(j\right)\right).\]
The decomposition in \eqref{eq:j*} is termed the Simple Transposition
Decomposition (STD) of $\sigma$. The cost of the STD of $\sigma$,
denoted by $S\left(\sigma\right)$, equals \begin{align*}
S\left(\sigma\right) & =\sum_{i\in\sigma}\varphi^{*}\left(i,\sigma\left(i\right)\right)-\varphi^{*}\left(j^{*},\sigma\left(j^{*}\right)\right).\end{align*}

\begin{thm}
\label{thm:MLS4M-1} For a cycle $\sigma$,
$M\left(\sigma\right)\le L\left(\sigma\right)\le S\left(\sigma\right)\le4M\left(\sigma\right).$\end{thm}
\begin{IEEEproof}
Clearly, $M\left(\sigma\right)\le L\left(\sigma\right)$. It is easy
to see that the STD is itself an MLD and, thus, $L\left(\sigma\right)\le S\left(\sigma\right)$.
For $S\left(\sigma\right)$, we have\begin{align}
S\left(\sigma\right) & =\sum_{i\in\sigma}\varphi^{*}\left(i,\sigma\left(i\right)\right)-\varphi^{*}\left(j^{*},\sigma\left(j^{*}\right)\right)\nonumber \\
 & \le\sum_{i\in\sigma}\varphi^{*}\left(i,\sigma\left(i\right)\right)\nonumber \\
 & \le2\sum_{i\in\sigma}\cost\left(p^{*}\left(i,\sigma\left(i\right)\right)\right)\label{eq:STD-Cost}\end{align}
where the last inequality follows from \eqref{eq:p-hat-p}. To complete
the proof, we need to show that $M\left(\sigma\right)\ge\frac{1}{2}\sum_{i}\cost\left(p^{*}\left(i,\sigma\left(i\right)\right)\right)$.
Since this result is of independent importance in our subsequent derivations,
we state it and prove it in Lemma \ref{lem:M-best-path-1}.
\end{IEEEproof}
In order to prove Lemma \ref{lem:M-best-path-1}, we first prove Lemma
\ref{lem:consistent-costs-1} and a corollary. 

Consider a transposition cost function $\varphi$ and a h-transposition
cost function $\psi$. Recall from Section \ref{sec:Notation-and-Definitions}
that\[
\htrans ab{\sigma^{-1}\left(a\right)}\htrans ba{\sigma^{-1}\left(b\right)}\sigma=\left(ab\right)\sigma.\]

\begin{lem}
\label{lem:consistent-costs-1}The minimum cost of an h-decomposition
of $\sigma$ is upper-bounded by the cost of the MCD of $\sigma$,
provided that $\varphi$ and $\psi$ are consistent.\end{lem}
\begin{IEEEproof}
We prove the lemma by showing that there exists an h-decomposition
of $\sigma$ with cost $M\left(\sigma\right)$. Suppose that the MCD
of $\sigma$ is $\tau=t_{m}t_{m-1}\cdots t_{1}$, where $t_{i}=\left(a_{i}b_{i}\right)$,
$1\le i\le m$, and where $m$ is the length of the MCD. Let the permutation
$t_{i}t_{i-1}\cdots t_{1}$ be denoted by $\sigma_{i}$. The cost
of the MCD is \[
M_{\varphi}\left(\sigma\right)=\sum_{i=1}^{m}\varphi\left(a_{i},b_{i}\right).\]
By replacing each transposition $t_{i}=\left(a_{i}b_{i}\right)$ in
$\tau$ by a corresponding pair of h-transpositions $\htrans{a_{i}}{b_{i}}{\sigma_{i-1}^{-1}\left(a_{i}\right)}\htrans{b_{i}}{a_{i}}{\sigma_{i-1}^{-1}\left(b_{i}\right)}$,
one can see that \begin{eqnarray*}
M\left(\sigma\right) & = & \sum_{i=1}^{m}\left(\psi\left(a_{i},b_{i}\right)+\psi\left(b_{i},a_{i}\right)\right),\end{eqnarray*}
since $\psi$ and $\varphi$ are consistent. Hence, the h-decomposition
\begin{multline*}
\htrans{b_{m}}{a_{m}}{\sigma_{m-1}^{-1}\left(b_{m}\right)}\htrans{a_{m}}{b_{m}}{\sigma_{m-1}^{-1}\left(a_{m}\right)}\cdots\\
\htrans{b_{1}}{a_{1}}{\sigma_{0}^{-1}\left(b_{1}\right)}\htrans{a_{1}}{b_{1}}{\sigma_{0}^{-1}\left(a_{1}\right)}\end{multline*}
 has cost $M\left(\sigma\right)$. In other words, decomposing each
transposition in an MCD into h-transpositions establishes the claimed
result.\end{IEEEproof}
\begin{cor}
\label{cor:maximin-1}For a fixed $\varphi$ and a cycle $\sigma$,
one has \[
M\left(\sigma\right)\ge\max_{\psi}\min_{H}C_{\psi}\left(H\right)\]
where the maximum is taken over all h-transposition costs $\psi$
consistent with $\varphi$, and the minimum is taken over all h-decompositions
$H$ of $\sigma$ with cost $C_{\psi}\left(H\right)$.\queede\end{cor}
\begin{lem}
\label{lem:M-best-path-1}It holds that $M\left(\sigma\right)\ge\frac{1}{2}\sum_{i}\cost\left(p^{*}\left(i,\sigma\left(i\right)\right)\right)$.\end{lem}
\begin{IEEEproof}
Define $\psi_{1/2}$ as\[
\psi_{1/2}\left(a,b\right)=\psi_{1/2}\left(b,a\right)=\varphi\left(a,b\right)/2.\]
It is clear that $\psi_{1/2}$ is consistent with $\varphi$. Hence,
we have

\begin{eqnarray*}
M\left(\sigma\right) & \ge & \max_{\psi}\min_{H}C_{\psi}\left(H\right)\\
 & \ge & \min_{H}C_{\psi_{1/2}}\left(H\right)\\
 & \stackrel{\mathtt{(\star)}}{=} & \sum_{i}\sum_{\left(ab\right)\in p^{*}\left(i,\sigma\left(i\right)\right)}\psi_{1/2}\left(a,b\right)\\
 & = & \frac{1}{2}\sum_{i}\cost\left(p^{*}\left(i,\sigma\left(i\right)\right)\right)\end{eqnarray*}
where $\mathtt{(\star)}$ follows from the fact that the minimum cost
h-decomposition uses the shortest path $p^{*}\left(i,\sigma\left(i\right)\right)$
between $i$ and $\sigma\left(i\right)$. In this case, $i$ becomes
the predecessor of $\sigma\left(i\right)$ through the following sequence
of h-transpositions:\[
\htrans{v_{m}}{\sigma\left(i\right)}i\cdots\htrans{v_{1}}{v_{2}}i\htrans i{v_{1}}i\]
where $p^{*}\left(i,\sigma\left(i\right)\right)=iv_{1}v_{2}\cdots v_{m}\sigma\left(i\right)$
is the shortest path between $i$ and $\sigma\left(i\right)$. 
\end{IEEEproof}
Observe that Theorem \ref{thm:MLS4M-1} asserts
that \emph{a minimum cost MLD never exceeds the cost of the corresponding
MCD by more than a factor of four}. Hence, a minimum cost MLD represents
a good approximation for an MCD, independent of the choice of the
cost function. On the other hand, STDs and their corresponding path
search algorithms are attractive alternatives to MLDs and dynamic
programs, due to the fact that they are particularly simple to implement.
\begin{example}
\label{exa:13-1}Consider the cycle $\sigma=\left(12345\right)$ and
the cost function $\varphi$, with $\varphi\left(2,4\right)=\varphi\left(2,5\right)=\varphi\left(3,5\right)=1$,
and $\varphi\left(i,j\right)=100$ for all remaining transposition.
First, observe that the costs are not reduced according to Alg. \ref{alg:opt-trans}.
Nevertheless, one can use the upper-bound for the transposition cost
in terms of the shortest paths defined in the proof of Lemma \ref{lem:M-best-path-1}.
In this case, one obtains \[
M\left(\sigma\right)\ge\frac{1}{2}\left(100+2+3+2+100\right)=103.5.\]
For example, the second term in the sum corresponds to a path going
from $2$ to $5$ and then from $5$ to $3$. The cost of this path
is two. 

Since $M\left(\sigma\right)$ has to be an integer, it follows that
$M\left(\sigma\right)\ge104$. 

The optimized cost function, $\varphi^{*}$, obtained from Alg. \ref{alg:opt-trans}
gives \[
\varphi^{*}\left(i,j\right)=\begin{cases}
1, & \quad\left(ij\right)\in\left\{ \left(25\right),\left(35\right),\left(24\right)\right\} \\
3, & \quad\left(ij\right)\in\left\{ \left(23\right),\left(45\right)\right\} \\
5, & \quad\left(ij\right)=\left(34\right)\\
100, & \quad\text{otherwise }\end{cases}\]

A minimum cost MLD can be computed using the dynamic program of Alg.
\ref{alg:opt-cycle}. One minimum cost MLD equals $\tau_{L}=\left(45\right)\left(35\right)\left(12\right)\left(25\right)$,
and has cost $L\left(\sigma\right)=105$. By substituting each of
the transposition in $\tau_{L}$ with their minimum cost transposition
decomposition, we obtain $\left(24\right)\left(25\right)\left(24\right)\left(35\right)\left(12\right)\left(25\right)$. 

It is easy to see that \[
\tau_{s}=\left(12\right)\left(23\right)\left(34\right)\left(45\right)\]
is the STD of $\sigma$ with cost $S\left(\sigma\right)=100+3+5+3=111$.

Hence, the inequality $M\left(\sigma\right)\le L\left(\sigma\right)\le S\left(\sigma\right)\le4M\left(\sigma\right)$
holds. Furthermore, note that $\sigma$ is an even cycle, and hence
must have an even number of transpositions in any of its decompositions.
This shows that $M\left(\sigma\right)=L\left(\sigma\right)=105$.\queede
\end{example}

\subsection{Metric-Path and Extended-Metric-Path Cost Functions}

We show next that for two non-trivial families of cost functions,
one can improve upon the bounds of Theorem \ref{thm:MLS4M-1}.
For metric-path cost functions, a minimum cost MLD is actually an
MCD, i.e., $L\left(\sigma\right)=M\left(\sigma\right)$. For extended-metric-path
costs, it holds that $L\left(\sigma\right)\le2M\left(\sigma\right)$. 

Note that metric-path costs are not the only cost functions which
admit MCDs of the form of MLDs -- another example includes star transposition
costs. For such costs, one has $\varphi\left(i,j\right)=\infty$ for
all $i,j$ except for one index $i$. The remaining costs are arbitrary,
but non-negative. The proof for this special case is straightforward
and hence omitted.
\begin{lem}
\label{lem:MLD-metric-tree}For a cycle $\sigma$ and a metric-path
cost function $\varphi$, $L\left(\sigma\right)\le\frac{1}{2}\sum_{i}\varphi\left(i,\sigma\left(i\right)\right)=\frac{1}{2}\sum_{i}\cost\left(p^{*}\left(i,\sigma\left(i\right)\right)\right)$. \end{lem}
\begin{IEEEproof}
The equality in the lemma follows from the definition of metric-path
cost functions. 

We recursively construct a spanning tree $T\left(\sigma\right)$ of
cost $B\left(\sigma\right)=\frac{1}{2}\sum_{i}\varphi\left(i,\sigma\left(i\right)\right)$,
such that $\mathcal{G}\left(\sigma\right)\cup T\left(\sigma\right)$
is planar. Since $T\left(\sigma\right)$ corresponds to an MLD, $L\left(\sigma\right)\le B\left(\sigma\right)$.

The validity of the recursive construction can be proved by induction.
For $k=2$, $T\left(\sigma\right)$ is the edge $\left(12\right)$.
Assume next that the cost of $T\left(\sigma\right)$ for any cycle
of length $\le k-1$ equals $B\left(\sigma\right)$.

For a cycle of length $k$, without loss of generality, assume that
the vertex labeled $1$ is a leaf in $\Theta_{s}$, the defining path
of $\varphi$, and that $t$ is its parent. We construct $T\left(\sigma\right)$
from smaller trees by letting \[
T\left(\sigma\right)=\left(1t\right)\cup T\left(\left(2\cdots t\right)\right)\cup T\left(\left(t\cdots k\right)\right).\]

See Figure \ref{fig:MLD-metric-tree} for an illustration. The cost
of $T\left(\sigma\right)$ is equal to $B\left(\left(2\cdots t\right)\right)+B\left(\left(t\cdots k\right)\right)+\varphi\left(1,t\right)$.
Note that we can write\begin{align*}
B\left(\left(2\cdots t\right)\right) & =\frac{1}{2}\sum_{i=2}^{t-1}\varphi\left(i,\sigma\left(i\right)\right)+\frac{1}{2}\varphi\left(2,t\right)\\
 & =\frac{1}{2}\sum_{i=1}^{t-1}\varphi\left(i,\sigma\left(i\right)\right)+\frac{1}{2}\varphi\left(2,t\right)-\frac{1}{2}\varphi\left(1,2\right),\end{align*}
 \begin{align*}
B\left(\left(t\cdots k\right)\right) & =\frac{1}{2}\sum_{i=t}^{k-1}\varphi\left(i,\sigma\left(i\right)\right)+\frac{1}{2}\varphi\left(t,k\right)\\
 & =\frac{1}{2}\sum_{i=t}^{k}\varphi\left(i,\sigma\left(i\right)\right)+\frac{1}{2}\varphi\left(t,k\right)-\frac{1}{2}\varphi\left(1,k\right).\end{align*}
Since $\varphi\left(1,2\right)=\varphi\left(1,t\right)+\varphi\left(t,2\right)$
and $\varphi\left(1,k\right)=\varphi\left(1,t\right)+\varphi\left(t,k\right)$,
it follows that\[
B\left(\left(2\cdots t\right)\right)+B\left(\left(t\cdots k\right)\right)=B\left(\left(1\cdots k\right)\right)-\varphi\left(1,t\right).\]
 This completes the proof of the Lemma.
\end{IEEEproof}
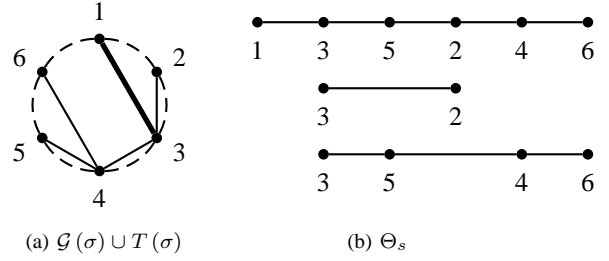
\begin{figure}
\subfloat[$\mathcal{G}\left(\sigma\right)\cup T\left(\sigma\right)$]{\centering
\psset{unit=25pt}
\begin{pspicture}[showgrid=false](-1.75,-1.75)(1.75,1.75)
\cnode*(0.000000,1.000000){2pt}{N1}

\nput{90}{N1}{1}

\cnode*(0.866025,0.500000){2pt}{N2}

\nput{30}{N2}{2}

\cnode*(0.866025,-0.500000){2pt}{N3}

\nput{-30}{N3}{3}

\cnode*(0.000000,-1.000000){2pt}{N4}

\nput{-90}{N4}{4}

\cnode*(-0.866025,-0.500000){2pt}{N5}

\nput{-150}{N5}{5}

\cnode*(-0.866025,0.500000){2pt}{N6}

\nput{150}{N6}{6}

\ncarc[arcangle=30,linestyle=dashed]{N1}{N2}

\ncarc[arcangle=30,linestyle=dashed]{N2}{N3}

\ncarc[arcangle=30,linestyle=dashed]{N3}{N4}

\ncarc[arcangle=30,linestyle=dashed]{N4}{N5}

\ncarc[arcangle=30,linestyle=dashed]{N5}{N6}

\ncarc[arcangle=30,linestyle=dashed]{N6}{N1}

\ncarc[arcangle=0,linestyle=solid,linewidth=2pt]{N1}{N3}

\ncarc[arcangle=0,linestyle=solid]{N2}{N3}

\ncarc[arcangle=0,linestyle=solid]{N3}{N4}

\ncarc[arcangle=0,linestyle=solid]{N4}{N6}

\ncarc[arcangle=0,linestyle=solid]{N5}{N4}

\end{pspicture}

}\subfloat[$\Theta_{s}$]{\centering
\psset{unit=25pt}
\begin{pspicture}[showgrid=false](-.50,-3)(4,1)
\cnode*(0,0){2pt}{N1}

\nput{-90}{N1}{1}

\cnode*(1,0){2pt}{N3}

\nput{-90}{N3}{3}

\ncarc[arcangle=0,linestyle=solid]{N1}{N3}

\cnode*(2,0){2pt}{N5}

\nput{-90}{N5}{5}

\ncarc[arcangle=0,linestyle=solid]{N3}{N5}

\cnode*(3,0){2pt}{N2}

\nput{-90}{N2}{2}

\ncarc[arcangle=0,linestyle=solid]{N5}{N2}

\cnode*(4,0){2pt}{N4}

\nput{-90}{N4}{4}

\ncarc[arcangle=0,linestyle=solid]{N2}{N4}

\cnode*(5,0){2pt}{N6}

\nput{-90}{N6}{6}

\ncarc[arcangle=0,linestyle=solid]{N4}{N6}

\cnode*(1,-1){2pt}{N32}

\nput{-90}{N32}{3}

\cnode*(3,-1){2pt}{N22}

\nput{-90}{N22}{2}

\ncarc[arcangle=0,linestyle=solid]{N22}{N32}

\cnode*(1,-2){2pt}{N32}

\nput{-90}{N32}{3}

\cnode*(2,-2){2pt}{N5}

\nput{-90}{N5}{5}

\cnode*(4,-2){2pt}{N4}

\nput{-90}{N4}{4}

\ncarc[arcangle=0,linestyle=solid]{N5}{N4}

\ncarc[arcangle=0,linestyle=solid]{N32}{N5}

\cnode*(5,-2){2pt}{N6}

\nput{-90}{N6}{6}

\ncarc[arcangle=0,linestyle=solid]{N4}{N6}

\end{pspicture}

}\caption{Example illustrating the proof of Lemma \ref{lem:MLD-metric-tree}:
(a) The cycle $\sigma=\left(12345\right)$. Edges of $\mathcal{G}\left(\sigma\right)$
are shown with dashed arcs. The edge $\left(13\right)$, shown with
a thick solid line, belongs to $T\left(\sigma\right)$. The tree $T\left(\left(3456\right)\right)$
consists of solid edges on the left hand side of $\left(13\right)$
and $T\left(\left(23\right)\right)$ consists of solid edges on the
right hand side of $\left(13\right)$. As stated in the proof, we
have $T\left(\sigma\right)=\left(13\right)\cup T\left(\left(23\right)\right)\cup T\left(\left(3456\right)\right)$.
(b) The defining path $\Theta_{s}$ of $\varphi$. Vertex $1$ is
a leaf and $3$ is its parent.}

\label{fig:MLD-metric-tree}
\end{figure}

\begin{thm}
\label{lem:MLD-construction-metric-tree}For a cycle $\sigma$ and
a metric-path cost function, one has\[
L\left(\sigma\right)=M\left(\sigma\right)=\frac{1}{2}\sum_{i}\varphi\left(i,\sigma\left(i\right)\right).\]
\end{thm}
\begin{IEEEproof}
Since $L\left(\sigma\right)\ge M\left(\sigma\right)$, it suffices
to show that $L\left(\sigma\right)\le\frac{1}{2}\sum_{i}\varphi\left(i,\sigma\left(i\right)\right)$
and $M\left(\sigma\right)\ge\frac{1}{2}\sum_{i}\varphi\left(i,\sigma\left(i\right)\right)$.
Lemma \ref{lem:MLD-metric-tree} establishes that $L\left(\sigma\right)\le\frac{1}{2}\sum_{i}\varphi\left(i,\sigma\left(i\right)\right)$.
From Lemma \ref{lem:M-best-path-1}, it also follows that \[
M\left(\sigma\right)\ge\frac{1}{2}\sum_{i}\cost\left(p^{*}\left(i,\sigma\left(i\right)\right)\right).\]
Since $\varphi$ is a metric-path cost function, we have $\varphi\left(i,\sigma\left(i\right)\right)=\cost\left(p^{*}\left(i,\sigma\left(i\right)\right)\right)$.
This proves the claimed result.\end{IEEEproof}
\begin{thm}
\label{thm:L<=00003D2M-ext-metric-tree}For extended-metric-path cost
functions $\varphi_{e}$, $L_{\varphi_{e}}\left(\sigma\right)\le2M_{\varphi_{e}}\left(\sigma\right)$.\end{thm}
\begin{IEEEproof}
We prove the theorem by establishing that \[
L_{\varphi_{e}}\left(\sigma\right)\stackrel{\mathsf{\left(a\right)}}{\le}\sum_{i}\cost\left(p^{*}\left(i,\sigma\left(i\right)\right)\right)\stackrel{\mathsf{\left(b\right)}}{\le}2M_{\varphi_{e}}\left(\sigma\right)\]
where $p^{*}\left(i,\sigma\left(i\right)\right)$ is the shortest
path between $i$ and $\sigma\left(i\right)$ in $\mathcal{K}\left(\varphi_{e}\right)$
and is calculated with respect to the cost function $\varphi_{e}$.

Let $\Theta_{s}$ be the defining path of an extended-metric-path
cost $\varphi_{e}$. Consider the \emph{metric-path cost} function,
$\varphi_{m}$, with defining path $\Theta_{s}$, and with costs of
all edges $\left(ij\right)\in\Theta_{s}$ doubled. If the edge $\left(ij\right)\notin\Theta_{s}$,
and if $c_{1}c_{2}\cdots c_{l+1}$ is the unique path from $c_{1}=i$
to $c_{l+1}=j$ in $\Theta_{s}$, then \[
\varphi_{m}\left(i,j\right)=\sum_{t=1}^{l}\varphi_{m}\left(c_{t},c_{t+1}\right)=2\sum_{t=1}^{l}\varphi_{e}\left(c_{t},c_{t+1}\right).\]
By \eqref{eq:def-navin}, $\varphi_{e}\left(i,j\right)\le\varphi_{m}\left(i,j\right)$,
for all $i,j$. Hence, $L_{\varphi_{e}}\left(\sigma\right)\le L_{\varphi_{m}}\left(\sigma\right)$.
Now, following along the same lines of the proof of Lemma \ref{lem:MLD-metric-tree},
it can be shown that \begin{align}
L_{\varphi_{e}}\left(\sigma\right) & \le L_{\varphi_{m}}\left(\sigma\right)\label{eq:lower-bound-non-metric}\\
 & =\frac{1}{2}\sum_{i}\varphi_{m}\left(i,\sigma\left(i\right)\right)\nonumber \\
 & =\sum_{i}\cost\left(p^{*}\left(i,\sigma\left(i\right)\right)\right),\nonumber \end{align}
which proves $\mathsf{\left(a\right)}$. 

Note that Lemma \ref{lem:M-best-path-1} holds for all non-negative
cost functions, including extended-metric-path cost functions. Thus,
\[
M_{\varphi_{e}}\left(\sigma\right)\ge\frac{1}{2}\sum_{i}\cost\left(p^{*}\left(i,\sigma\left(i\right)\right)\right),\]
 which proves $\mathsf{\left(b\right)}$.\end{IEEEproof}
\begin{example}
Consider the cycle $\sigma=\left(12345\right)$ and the extended-metric-path
cost function of Example \ref{exa:navin}.

By inspection, one can see that an MCD of $\sigma$ is $\left(14\right)\left(13\right)\left(35\right)\left(24\right)\left(14\right)\left(13\right)$,
with cost $M\left(\sigma\right)=6$. A minimum cost MLD of $\sigma$
is $\left(14\right)\left(23\right)\left(13\right)\left(45\right)$,
with cost $L\left(\sigma\right)=8$. The STD is $\left(12\right)\left(23\right)\left(34\right)\left(45\right)$,
with cost $S\left(\sigma\right)=12$. Thus, we observe that the inequality
$L\left(\sigma\right)\le S\left(\sigma\right)\le2M\left(\sigma\right)$
is satisfied.\queede
\end{example}
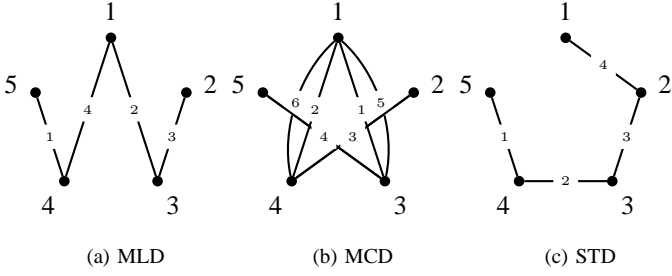
\begin{figure}
\subfloat[MLD]{\centering
\psset{unit=30pt}
\begin{pspicture}[showgrid=false](-1.25,-1.5)(1.5,1.5)
\cnode*(0.000000,1.000000){2pt}{N1}

\nput{90}{N1}{1}

\cnode*(0.951057,0.309017){2pt}{N2}

\nput{18}{N2}{2}

\cnode*(0.587785,-0.809017){2pt}{N3}

\nput{-54}{N3}{3}

\cnode*(-0.587785,-0.809017){2pt}{N4}

\nput{-126}{N4}{4}

\cnode*(-0.951057,0.309017){2pt}{N5}

\nput{-198}{N5}{5}

\ncline{-}{N1}{N3}

\ncput*{\tiny $2$}

\ncline{-}{N1}{N4}

\ncput*{\tiny $4$}

\ncline{-}{N2}{N3}

\ncput*{\tiny $3$}

\ncline{-}{N4}{N5}

\ncput*{\tiny $1$}

\end{pspicture}

}\subfloat[MCD]{\centering
\psset{unit=30pt}
\begin{pspicture}[showgrid=false](-1.25,-1.5)(1.5,1.5)
\cnode*(0.000000,1.000000){2pt}{N1}

\nput{90}{N1}{1}

\cnode*(0.951057,0.309017){2pt}{N2}

\nput{18}{N2}{2}

\cnode*(0.587785,-0.809017){2pt}{N3}

\nput{-54}{N3}{3}

\cnode*(-0.587785,-0.809017){2pt}{N4}

\nput{-126}{N4}{4}

\cnode*(-0.951057,0.309017){2pt}{N5}

\nput{-198}{N5}{5}

\ncline{-}{N1}{N3}

\ncput*{\tiny $1$}

\ncline{-}{N1}{N4}

\ncput*{\tiny $2$}

\ncline{-}{N2}{N4}

\ncput*{\tiny $3$}

\ncline{-}{N3}{N5}

\ncput*{\tiny $4$}

\ncarc[arcangle=30]{-}{N1}{N3}

\ncput*{\tiny $5$}

\ncarc[arcangle=-30]{-}{N1}{N4}

\ncput*{\tiny $6$}

\ncline{-}{N2}{N4}

\ncput*{\tiny $3$}

\ncline{-}{N3}{N5}

\ncput*{\tiny $4$}

\end{pspicture}

}\subfloat[STD]{\centering
\psset{unit=30pt}
\begin{pspicture}[showgrid=false](-1.25,-1.5)(1.5,1.5)
\cnode*(0.000000,1.000000){2pt}{N1}

\nput{90}{N1}{1}

\cnode*(0.951057,0.309017){2pt}{N2}

\nput{18}{N2}{2}

\cnode*(0.587785,-0.809017){2pt}{N3}

\nput{-54}{N3}{3}

\cnode*(-0.587785,-0.809017){2pt}{N4}

\nput{-126}{N4}{4}

\cnode*(-0.951057,0.309017){2pt}{N5}

\nput{-198}{N5}{5}

\ncline{-}{N4}{N5}

\ncput*{\tiny $1$}

\ncline{-}{N3}{N4}

\ncput*{\tiny $2$}

\ncline{-}{N2}{N3}

\ncput*{\tiny $3$}

\ncline{-}{N1}{N2}

\ncput*{\tiny $4$}

\end{pspicture}

}

\caption{MLD (a) and MCD (b) for $\sigma=\left(12345\right)$. Edge labels
denote the order in which transpositions are applied\label{fig:ex-conter-ex}}

\end{figure}

\section{Optimizing Permutations with Multiple Cycles}

Most of the results in the previous section generalize to permutations
with multiple cycles without much difficulty. We present next the
generalization of those results.

Let $\pi$ be a permutation in $\mathbb{S}_{n}$, with cycle decomposition
$\sigma_{1}\sigma_{2}\cdots\sigma_{\ell}$. A decomposition of $\pi$
with minimum number of transpositions is the product of MLDs of individual
cycles $\sigma_{i}$. Thus, the minimum cost MLD of $\pi$ equals
\[
L\left(\pi\right)=\sum_{t=1}^{\ell}L\left(\sigma_{t}\right).\]
The STD of $\pi$ is the product of the STDs of individual cycles
$\sigma_{i}$.

The following theorem generalizes the results presented for single
cycle permutations to permutations with multiple cycles.
\begin{thm}
Consider a permutation $\pi$ with cycle decomposition $\sigma_{1}\sigma_{2}\cdots\sigma_{\ell}$,
and cost function $\varphi$. The following claims hold.
\begin{enumerate}
\item $S\left(\pi\right)\le2\sum_{i}\cost\left(p^{*}\left(i,\pi\left(i\right)\right)\right).$
\item $M\left(\pi\right)\ge\frac{1}{2}\sum_{i}\cost\left(p^{*}\left(i,\pi\left(i\right)\right)\right).$
\item \textup{$L\left(\pi\right)\le S\left(\pi\right)\le4M\left(\pi\right).$}
\item If $\varphi$ is a metric-path cost function, then \[
M\left(\pi\right)=L\left(\pi\right).\]

\item If $\varphi$ is an extended-metric-path cost function, then \[
L\left(\pi\right)\le2M\left(\pi\right).\]

\end{enumerate}
\end{thm}
\begin{IEEEproof}
~
\begin{enumerate}
\item For each cycle it holds that \[
S\left(\sigma_{t}\right)\le2\sum_{i\in\sigma_{t}}\cost\left(p^{*}\left(i,\sigma_{t}\left(i\right)\right)\right),\]
which can be seen by referring to \eqref{eq:STD-Cost} in the proof
of Theorem \ref{thm:MLS4M-1}. Thus, \begin{align*}
S\left(\pi\right) & =\sum_{t=1}^{\ell}S\left(\sigma_{t}\right)\\
 & \le\sum_{t=1}^{\ell}2\sum_{i\in\sigma_{t}}\cost\left(p^{*}\left(i,\pi\left(i\right)\right)\right)\\
 & =2\sum_{i=1}^{n}\cost\left(p^{*}\left(i,\pi\left(i\right)\right)\right).\end{align*}

\item The same argument as in Lemma \ref{lem:M-best-path-1} applies without
modifications.
\item For each $1\le t\le\ell$, from the proof of Theorem \ref{thm:MLS4M-1},
we have $L\left(\sigma_{t}\right)\le S\left(\sigma_{t}\right)$. Consequently,\[
L\left(\pi\right)=\sum_{t=1}^{\ell}L\left(\sigma_{t}\right)\le\sum_{t=1}^{\ell}S\left(\sigma_{t}\right)=S\left(\pi\right).\]
Furthermore, from parts 1 and 2 of this theorem, it follows that $S\left(\pi\right)\le4M\left(\pi\right)$.
Therefore $L\left(\pi\right)\le S\left(\pi\right)\le4M\left(\pi\right)$.
\item From Lemma \ref{lem:MLD-metric-tree}, for each $\sigma_{t}$, it
holds that \[
L\left(\sigma_{t}\right)\le\frac{1}{2}\sum_{i\in\sigma_{t}}\cost\left(p^{*}\left(i,\pi\left(i\right)\right)\right).\]
By summing over all cycles, we obtain \[
L\left(\pi\right)\le\frac{1}{2}\sum_{i=1}^{n}\cost\left(p^{*}\left(i,\pi\left(i\right)\right)\right).\]
The claimed result follows from part 2 and the fact that $M\left(\pi\right)\le L\left(\pi\right)$.
\item From the proof of Theorem \ref{thm:L<=00003D2M-ext-metric-tree},
we have $L\left(\sigma_{t}\right)\le\sum_{i\in\sigma_{t}}\cost\left(p^{*}\left(i,\pi\left(i\right)\right)\right)$.
By summing over all cycles, we obtain\[
L\left(\pi\right)\le\sum_{i=1}^{n}\cost\left(p^{*}\left(i,\pi\left(i\right)\right)\right)\le2M\left(\pi\right),\]
where the last inequality follows from part 2 of this theorem.
\end{enumerate}
\end{IEEEproof}

\subsection{Merging cycles}

In Section \ref{sec:Minimum-Cost-MLDs}, we demonstrated that the
minimum cost of an MLD for an arbitrary permutation represents a constant
approximation for an MCD. The MLD of a permutation represents the
product of the MLDs of individual cycles of the permutation. Clearly,
optimization of individual cycle costs may not lead to the minimum
cost decomposition of a permutation. For example, it may happen that
the cost of transpositions within a cycle are much higher than the
costs of transpositions between elements in different cycles. It is
therefore useful to analyze how merging of cycles may affect the overall
cost of a decomposition. 

We propose a simple merging method that consists of two steps:
\begin{enumerate}
\item Find a sequence of transpositions \[
\tau'=t_{k-1}\cdots t_{1}\]
so that $\sigma'=\tau'\pi$ is a single cycle. Ideally, this sequence
should have minimum cost, although this is not required in the proofs
to follow. 
\item Find the minimum cost MLD $\tau$ of $\sigma'$.
\end{enumerate}
The resulting decomposition is of the form $\tau'^{-1}\tau$.

Suppose that $\pi$ has $k$ cycles. Joining $k$ cycles requires
$k-1$ transpositions. Hence, each $t_{i}$ is a transposition joining
two cycles of $\pi$. The cost of $\tau'$ equals $\sum_{i=1}^{k-1}\varphi\left(a_{i},b_{i}\right)$,
where $t_{i}=\left(a_{i}b_{i}\right)$. The cost of the resulting
decomposition using $\tau'$ and the single cycle MLD equals \begin{equation}
C=\sum_{i=1}^{k-1}\varphi\left(a_{i},b_{i}\right)+L(\sigma').\label{eq:app1-1}\end{equation}
 Since $\pi=t_{1}\cdots t_{k}\sigma'$, we also have \begin{equation}
L(\sigma')\le4M\left(\sigma'\right)\le4\left(\sum_{i=1}^{k-1}\varphi\left(a_{i},b_{i}\right)+M(\pi)\right).\label{eq:app2-1}\end{equation}
 Hence, from \eqref{eq:app1-1} and \eqref{eq:app2-1}, $C$ is upper
bounded by \begin{equation}
\begin{split}C & \le\sum_{i=1}^{k-1}\varphi\left(a_{i},b_{i}\right)+4\left(\sum_{i=1}^{k-1}\varphi\left(a_{i},b_{i}\right)+M(\pi)\right)\\
 & \le5k\varphi_{max}+4M(\pi),\end{split}
\end{equation}
where $\varphi_{max}$ is the highest cost in $\varphi$. The approximation
ratio, defined as $C/M(\pi)$, is upper bounded by \begin{equation}
\alpha\le4+\frac{5k}{n-k}\frac{\varphi_{max}}{\varphi_{min}}=4+\frac{5k/n}{1-k/n}\frac{\varphi_{max}}{\varphi_{min}},\label{eq:alpha}\end{equation}
 which follows from the fact $M(\pi)\ge\left(n-k\right)\varphi_{min}$,
where $\varphi_{min}$ is the smallest cost in $\varphi$, assumed
to be nonzero.

Although $\alpha$ is bounded by a value strictly larger than four,
according to the expression above, this does not necessarily imply
that merging cycles is sub-optimal compared to running the MLD algorithm
on individual cycles. Furthermore, if the MCDs of single cycles can
be computed correctly, one can show that \begin{eqnarray}
C & = & \sum_{i=1}^{k-1}\varphi\left(a_{i},b_{i}\right)+M(\sigma').\label{eq:app1-1-1}\\
 & \le2 & \sum_{i=1}^{k-1}\varphi\left(a_{i},b_{i}\right)+M(\pi)\\
 & \le & 2k\varphi_{max}+M\left(\pi\right)\end{eqnarray}
 The approximation ratio in this case is upper bounded by \[
\alpha\le1+\frac{2k}{n-k}\frac{\varphi_{max}}{\varphi_{min}}=1+\frac{2k/n}{1-k/n}\frac{\varphi_{max}}{\varphi_{min}}\cdot\]

\begin{lem}
\label{lem:perm-approx-1}Let $\pi$ be a randomly chosen permutation
from $\mathbb{S}_{n}$. Given that the MCDs of single cycles can be
computed correctly, and provided that $\varphi_{max}=o\left(n/\log n\right)$,
$\alpha$ goes to one in probability as $n\rightarrow\infty$.\end{lem}
\begin{IEEEproof}
Let $X_{n}$ be the random variable denoting the number of cycles
in a random permutation $\pi_{n}\in\mathbb{S}_{n}$. It is well known
that $EX_{n}=\sum_{j=1}^{n}\frac{1}{j}=H\left(n\right)$ and that
$EX_{n}\left(X_{n}-1\right)=\left(EX_{n}\right)^{2}-\sum_{j=1}^{n}\frac{1}{j^{2}}$
\cite{knuth2005art}. Here, $H\left(n\right)$ denotes the $n$th
Harmonic number. Thus \begin{equation}
\begin{split}EX_{n}^{2} & =O\left(\left(\ln n\right)^{2}\right)\end{split}
,\end{equation}
 which shows that $X_{n}/n\to0$ in quadratic mean as $n\to\infty$.
Hence $X_{n}/n\to0$ in probability. By Slutsky's theorem \cite{slutsky},
$\alpha\to1$ in probability as $n\to\infty$. 
\end{IEEEproof}
In the following example, all operations are performed modulo 10,
with zero replaced by 10. 
\begin{example}
Consider the permutation $\pi=\sigma_{1}\sigma_{2}$, where $\sigma_{1}=\left(1\,7\,3\,9\,5\right)$
and $\sigma_{2}=\left(2\,8\,4\,10\,6\right)$, and the cost function
$\varphi$,\[
\varphi\left(i,j\right)=\begin{cases}
1, & \quad d\left(i,j\right)=1\\
\infty, & \quad\text{otherwise}\end{cases}\]
where $d\left(i,j\right)=\min\left\{ \left|i-j\right|,10-\left|i-j\right|\right\} $.
Note that \[
\cost\left(p^{*}\left(i,j\right)\right)=d\left(i,j\right).\]
where $p^{*}$ is the shortest path from $i$ to $j$. We make the
following observations regarding the decompositions of $\pi$.
\begin{enumerate}
\item MCD: We cannot find the MCD of $\pi$, but we can easily obtain the
following bound: \[
M\left(\pi\right)\ge\lceil\frac{1}{2}\sum_{i=1}^{10}\cost\left(p^{*}\left(i,\pi\left(i\right)\right)\right)\rceil=\frac{2\cdot5\cdot4}{2}=20.\]
 
\item MLD: As before, let the output of Alg. \ref{alg:opt-trans} be denoted
by $\varphi^{*}$. We have $\varphi^{*}\left(i,j\right)=2d\left(i,j\right)-1$.
The minimum cost MLDs for the cycles are\begin{eqnarray*}
\left(1\,7\,3\,9\,5\right) & = & \left(1\,9\right)\left(3\,7\right)\left(1\,3\right)\left(9\,5\right),\\
\left(2\,8\,4\,10\,6\right) & = & \left(2\,10\right)\left(4\,8\right)\left(2\,4\right)\left(6\,10\right),\end{eqnarray*}
each of cost 20. The MLD of $\pi$ is the concatenation of the MLDs
of $\sigma_{1}$ and $\sigma_{2}$: \[
\pi=\left(2\,10\right)\left(4\,8\right)\left(2\,4\right)\left(6\,10\right)\left(1\,9\right)\left(3\,7\right)\left(1\,3\right)\left(9\,5\right),\]
with overall cost equal to 40. 
\item STD: It can be shown that the STD of $\sigma_{1}$ is\begin{eqnarray*}
\sigma_{1} & = & \left(1\ 7\right)\left(7\ 3\right)\left(3\ 9\right)\left(9\ 5\right),\\
\sigma_{2} & = & \left(2\ 8\right)\left(8\ 4\right)\left(4\ 10\right)\left(10\ 6\right),\end{eqnarray*}
each with cost 28. The total cost of the STD is $S\left(\pi\right)=56$.
\item Merging cycles: Instead of finding the minimum cost MLD of each cycle
separately, we may join the cycles and find the MLD of a larger cycle.
Here, we find the MLD of $\sigma'=\left(1\,2\right)\pi=\left(1\,7\,3\,9\,5\,2\,8\,4\,10\,6\right)$.
The cost of the minimum MLD of $\sigma'$ can be shown to be 37. Since
the cost of the transposition $\left(1\,2\right)$ must also be accounted
for, the total cost is 38. Observe that this cost is smaller than
the MLD cost of part 2, and hence merging cycles may provide better
solutions than the ones indicated by the bound \eqref{eq:alpha} or
as obtained through optimization of individual cycles.\queede
\end{enumerate}
\end{example}

\section{Conclusions\label{sec:Conclusions}}

We introduced the problem of minimum cost transposition sorting and
presented an algorithm for computing a transposition decomposition
of an arbitrary permutation, with cost at most four times the minimum
cost. We also described an algorithm that finds the minimum cost of
each transposition in terms of a product of other transpositions,
as well as an algorithm that computes the minimum cost/minimum length
decomposition using dynamic programing methods.

We also showed that more accurate solutions are possible for two particular
families of cost functions: for metric-path costs, we derived optimal
decomposition algorithms, while for extended-metric-path costs, we
described a 2-approximation method. 

The algorithms presented in this paper are of polynomial complexity.
Finding the minimum cost of a transposition has complexity $O(n^{4})$.
Given the optimized cost transpositions, the minimum length decomposition
can also be constructed in $O(n^{4})$ steps. Computing a decomposition
whose cost does not exceed the minimum cost by more than a factor
of four requires $O(n^{4})$ steps as well. 
\begin{acknowledgement*}
The authors gratefully acknowledge useful discussions with Chien-Yu
Chen, Chandra Chekuri, and Alon Orlitsky. They would also like to
thank Navin Kashyap for describing Example 1. This work was funded
by the NSF grants NSF CCF 08-21910 and NSF CCF 08-09895.
\end{acknowledgement*}
\appendix

\section{The Bellman-Ford Algorithm}

We describe an algorithm for finding $\bar{\varphi}\left(\hat{p}\left(a,b\right)\right)$.
The algorithm represents a variant of the Bellman-Ford procedure,
described in detail in \cite{cormen24introduction}.

Recall that for each path $p=v_{1}\cdots v_{m+1}$ in $\mathcal{K}\left(\varphi\right)$,
we define two types of costs: the standard cost of the path, \begin{equation}
\cost\left(p\right)=\sum_{i=1}^{m}\varphi\left(v_{i},v_{i+1}\right),\label{eq:cost}\end{equation}
and the transposition path cost, \begin{equation}
\bar{\varphi}\left(p\right)=2\cost\left(p\right)-\max_{i}\varphi\left(v_{i},v_{i+1}\right).\label{eq:barvarphi}\end{equation}
The goal is to find the path that minimizes the transposition path
cost in \eqref{eq:barvarphi}. 

Before describing our algorithm, we briefly review the standard Single-Source
Bellman-Ford shortest path algorithm, and its relaxation techniques. 

Given a fixed source $s$, for each vertex $v\neq s$, the algorithm
maintains an upper bound on the distance between $s$ and $v$, denoted
by $D\left(v\right)$. Initially, for each vertex $v$, we have $D\left(v\right)=\varphi(s,v)$. 

{}``Relaxing'' an edge $\left(uv\right)$ means testing that the
upper-bounds $D\left(u\right)$ and $D\left(v\right)$ satisfy the
conditions,\begin{align}
D\left(u\right) & \le D\left(v\right)+w,\nonumber \\
D\left(v\right) & \le D\left(u\right)+w,\label{eq:relaxation}\end{align}
where $w$ denotes the cost of the edge $(uv)$. If the above conditions
are not satisfied, then one of the two upper-bounds can be improved,
since one can reach $u$ by passing through $v$, and vice versa.

In our algorithm, we maintain the upper-bound for two types of costs.
The source $s$ is an arbitrary vertex in $\mathcal{K}(\varphi)$.
For a path between $s$ and a vertex $v$, we use $D_{1}\left(v\right)$
to denote the bound on the minimum transposition path cost, and we
use $D_{2}\left(v\right)$ to denote the bound on twice the minimum
cost of the path. From the definitions of these costs, it is clear
that\begin{align}
D_{2}\left(u\right) & \le2w+D_{2}\left(v\right),\label{eq:confusing}\\
D_{2}\left(v\right) & \le2w+D_{2}\left(u\right),\nonumber \\
D_{1}\left(u\right) & \le\min\left\{ w+D_{2}\left(v\right),2w+D_{1}\left(v\right)\right\} ,\nonumber \\
D_{1}\left(v\right) & \le\min\left\{ w+D_{2}\left(u\right),2w+D_{1}\left(u\right)\right\} .\nonumber \end{align}
The relaxation algorithm for these inequalities, Alg. \ref{alg:relax},
is straightforward to implement. To describe the properties of the
output of the Bellman-Ford algorithm, we briefly comment on a simple
property of the algorithm, termed the path-relaxation property.

Suppose $p=v_{1}\cdots v_{m+1}$ is the shortest path (in terms of
\eqref{eq:cost} or \eqref{eq:barvarphi}) from $s=v_{1}$ to $u=v_{m+1}$.
After relaxing the edges $\left(v_{1}v_{2}\right),\left(v_{2}v_{3}\right),\cdots\left(v_{m}v_{m+1}\right)$,
in that given order, the upper-bound $D_{i}\left(u\right)$ (for $i=1,2$)
equals the optimal cost of the corresponding path. Note that the property
still holds even if the relaxations of the edges $\left(v_{1}v_{2}\right),\left(v_{2}v_{3}\right),\cdots,\left(v_{m}v_{m+1}\right)$
are interleaved by relaxations of some other edges. In other words,
it suffices to identify only a subsequence of relaxations of the edges
$\left(v_{1}v_{2}\right),\left(v_{2}v_{3}\right),\cdots\left(v_{m}v_{m+1}\right)$. 

In the algorithm below, we use $pred_{i}(v)$ to denote the predecessor
of node $v$ used for tracking the updates of the cost $D_{i}(v)$,
$i=1,2$, and $(u,i),i=1,2$, to indicate from which of the two costs,
minimized over in \eqref{eq:confusing}, $u$ originated. Note that
this notion of predecessor is not to be confused with the predecessor
element in a two-line permutation representation.

\begin{algorithm}
\caption{\textsc{Relax$\left(u,v\right)$}}
\label{alg:relax}\begin{algorithmic}[1]
\State $w\leftarrow \varphi(u,v)$
\vspace{.2cm}
\If {$D_2(v)>D_2(u)+2w$}
    \State $D_2(v)\leftarrow D_2(u)+2w$
    \State $pred_2(v)\leftarrow (u,2)$
\EndIf
\vspace{.2cm}
\If {$D_2(u)>D_2(v)+2w$}
    \State $D_2(u)\leftarrow D_2(v)+2w$
    \State $pred_2(u)\leftarrow (v,2)$
\EndIf
\vspace{.2cm}
\If {$D_1(v)>D_2(u)+w$}
    \State $D_1(v)\leftarrow D_2(u)+w$
    \State $pred_1(v)\leftarrow (u,2)$
\EndIf
\vspace{.2cm}
\If {$D_1(u)>D_2(v)+w$}
    \State $D_1(u)\leftarrow D_2(v)+w$
    \State $pred(u,1)\leftarrow (v,2)$
\EndIf
\vspace{.2cm}
\If {$D_1(v)>D_1(u)+2w$}
    \State $D_1(v)\leftarrow D_1(u)+2w$
    \State $pred_1(v)\leftarrow (u,1)$
\EndIf
\vspace{.2cm}
\If {$D_1(u)>D_1(v)+2w$}
    \State $D_1(u)\leftarrow D_1(v)+2w$
    \State $pred_1(u)\leftarrow (v,1)$
\EndIf
\end{algorithmic}
\end{algorithm}

The Bellman-Ford algorithm performs $n-1$ rounds of relaxation on
the edges of the graph $\mathcal{K}(\varphi)$. Lemma \ref{lem:BF}
proves the correctness of the algorithm.

\begin{algorithm}
\begin{algorithmic}[1]
\State Input: vertex $s$
\State Output: $\hat{p}(s,u)$ for $1\le u\le n$
\For {$u\leftarrow 1\cdots n$}
   \State $D_1(u)\leftarrow \varphi(s,u)$
   \State $pred_1(u)\leftarrow s$
   \State $D_2(u)\leftarrow 2\varphi(s,u)$
   \State $pred_2(u)\leftarrow s$
\EndFor
\For {$i\leftarrow 1\cdots n-1$}
   \For {each edge $(uv)\in E(\mathcal{K}(\varphi))$}
      \State {\sc Relax}$(u,v)$
   \EndFor
\EndFor
\For {$u\leftarrow 1\cdots n$}
   \State initialize path at $u$
   \State backtrack min cost alg to recover path to $s$ 
   \State output $\hat{p}(s,u)$
\EndFor
\end{algorithmic}

\caption{\textsc{Single-Source Bellman-Ford$\left(s\right)$}}
\label{alg:SSBF}
\end{algorithm}

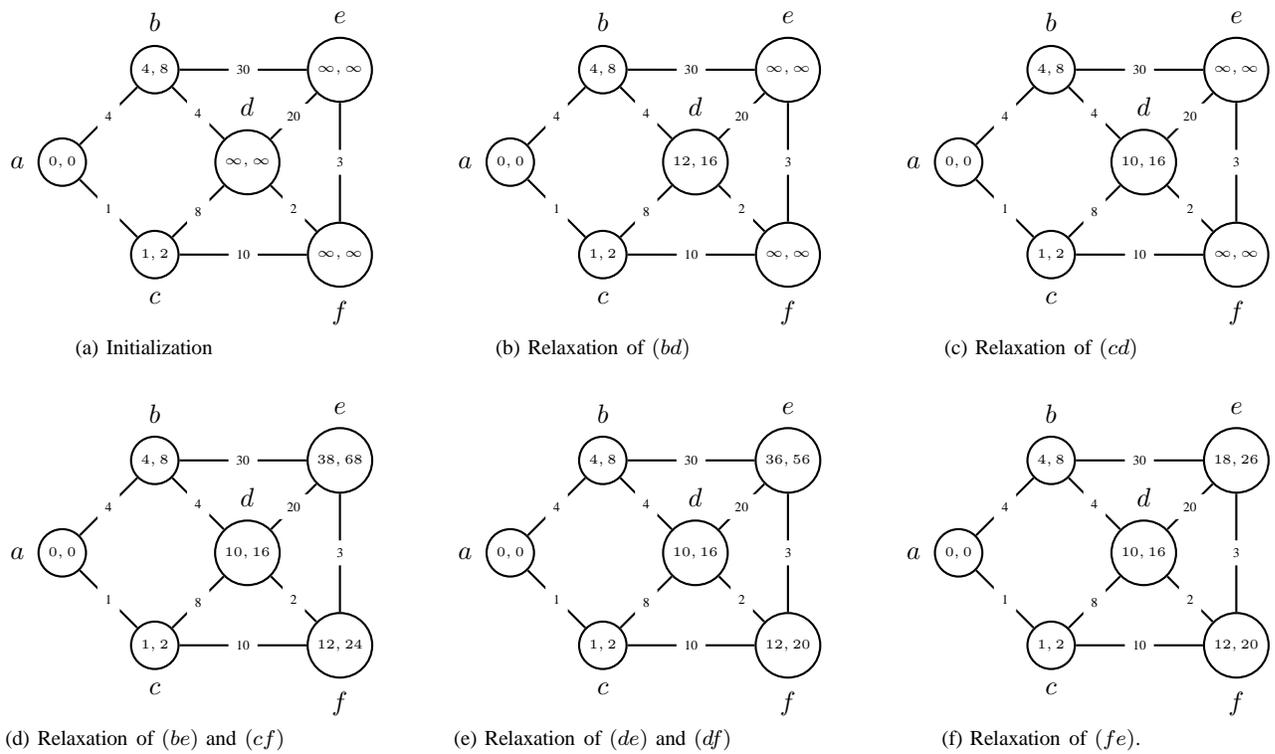
\begin{figure*}
\subfloat[Initialization]{
\centering

\psset{unit=50pt}

\begin{pspicture}[showgrid=false](-1,-1.25)(2.25,1.25)

\cnodeput{0}(0,0){a}{{\tiny $0,0$}}

\nput{180}{a}{$a$}

\cnodeput{0}(.7,.7){b}{{\tiny $4,8$}}

\nput{90}{b}{$b$}

\cnodeput{0}(0.7,-0.7){c}{{\tiny $1,2$}}

\nput{-90}{c}{$c$}

\cnodeput{0}(1.4,0){d}{{\tiny $\infty,\infty$}}

\nput{90}{d}{$d$}

\cnodeput{0}(2.1,0.7){e}{{\tiny $\infty,\infty$}}

\nput{90}{e}{$e$}

\cnodeput{0}(2.1,-0.7){f}{{\tiny $\infty,\infty$}}

\nput{-90}{f}{$f$}

\ncarc[arcangle=0]{a}{b}

\ncput*{\tiny 4}

\ncarc[arcangle=0]{a}{c}

\ncput*{\tiny 1}

\ncarc[arcangle=0]{d}{b}

\ncput*{\tiny 4}

\ncarc[arcangle=0]{d}{c}

\ncput*{\tiny 8}

\ncarc[arcangle=0]{d}{e}

\ncput*{\tiny 20}

\ncarc[arcangle=0]{d}{f}

\ncput*{\tiny 2}

\ncarc[arcangle=0]{b}{e}

\ncput*{\tiny 30}

\ncarc[arcangle=0]{c}{f}

\ncput*{\tiny 10}

\ncarc[arcangle=0]{f}{e}

\ncput*{\tiny 3}

\end{pspicture}
\label{fig:init}}
\subfloat[Relaxation of $\left(bd\right)$]{
\centering

\psset{unit=50pt}

\begin{pspicture}[showgrid=false](-1,-1.25)(2.25,1.25)

\cnodeput{0}(0,0){a}{{\tiny $0,0$}}

\nput{180}{a}{$a$}

\cnodeput{0}(.7,.7){b}{{\tiny $4,8$}}

\nput{90}{b}{$b$}

\cnodeput{0}(0.7,-0.7){c}{{\tiny $1,2$}}

\nput{-90}{c}{$c$}

\cnodeput{0}(1.4,0){d}{{\tiny $12,16$}}

\nput{90}{d}{$d$}

\cnodeput{0}(2.1,0.7){e}{{\tiny $\infty,\infty$}}

\nput{90}{e}{$e$}

\cnodeput{0}(2.1,-0.7){f}{{\tiny $\infty,\infty$}}

\nput{-90}{f}{$f$}

\ncarc[arcangle=0]{a}{b}

\ncput*{\tiny 4}

\ncarc[arcangle=0]{a}{c}

\ncput*{\tiny 1}

\ncarc[arcangle=0]{d}{b}

\ncput*{\tiny 4}

\ncarc[arcangle=0]{d}{c}

\ncput*{\tiny 8}

\ncarc[arcangle=0]{d}{e}

\ncput*{\tiny 20}

\ncarc[arcangle=0]{d}{f}

\ncput*{\tiny 2}

\ncarc[arcangle=0]{b}{e}

\ncput*{\tiny 30}

\ncarc[arcangle=0]{c}{f}

\ncput*{\tiny 10}

\ncarc[arcangle=0]{f}{e}

\ncput*{\tiny 3}

\end{pspicture}
\label{fig:bd}}
\subfloat[Relaxation of $\left(cd\right)$]{
\centering

\psset{unit=50pt}

\begin{pspicture}[showgrid=false](-1,-1.25)(2.25,1.25)

\cnodeput{0}(0,0){a}{{\tiny $0,0$}}

\nput{180}{a}{$a$}

\cnodeput{0}(.7,.7){b}{{\tiny $4,8$}}

\nput{90}{b}{$b$}

\cnodeput{0}(0.7,-0.7){c}{{\tiny $1,2$}}

\nput{-90}{c}{$c$}

\cnodeput{0}(1.4,0){d}{{\tiny $10,16$}}

\nput{90}{d}{$d$}

\cnodeput{0}(2.1,0.7){e}{{\tiny $\infty,\infty$}}

\nput{90}{e}{$e$}

\cnodeput{0}(2.1,-0.7){f}{{\tiny $\infty,\infty$}}

\nput{-90}{f}{$f$}

\ncarc[arcangle=0]{a}{b}

\ncput*{\tiny 4}

\ncarc[arcangle=0]{a}{c}

\ncput*{\tiny 1}

\ncarc[arcangle=0]{d}{b}

\ncput*{\tiny 4}

\ncarc[arcangle=0]{d}{c}

\ncput*{\tiny 8}

\ncarc[arcangle=0]{d}{e}

\ncput*{\tiny 20}

\ncarc[arcangle=0]{d}{f}

\ncput*{\tiny 2}

\ncarc[arcangle=0]{b}{e}

\ncput*{\tiny 30}

\ncarc[arcangle=0]{c}{f}

\ncput*{\tiny 10}

\ncarc[arcangle=0]{f}{e}

\ncput*{\tiny 3}

\end{pspicture}
\label{fig:cd}}

\subfloat[Relaxation of $\left(be\right)$ and $\left(cf\right)$]{
\centering

\psset{unit=50pt}

\begin{pspicture}[showgrid=false](-1,-1.25)(2.25,1.25)

\cnodeput{0}(0,0){a}{{\tiny $0,0$}}

\nput{180}{a}{$a$}

\cnodeput{0}(.7,.7){b}{{\tiny $4,8$}}

\nput{90}{b}{$b$}

\cnodeput{0}(0.7,-0.7){c}{{\tiny $1,2$}}

\nput{-90}{c}{$c$}

\cnodeput{0}(1.4,0){d}{{\tiny $10,16$}}

\nput{90}{d}{$d$}

\cnodeput{0}(2.1,0.7){e}{{\tiny $38,68$}}

\nput{90}{e}{$e$}

\cnodeput{0}(2.1,-0.7){f}{{\tiny $12,24$}}

\nput{-90}{f}{$f$}

\ncarc[arcangle=0]{a}{b}

\ncput*{\tiny 4}

\ncarc[arcangle=0]{a}{c}

\ncput*{\tiny 1}

\ncarc[arcangle=0]{d}{b}

\ncput*{\tiny 4}

\ncarc[arcangle=0]{d}{c}

\ncput*{\tiny 8}

\ncarc[arcangle=0]{d}{e}

\ncput*{\tiny 20}

\ncarc[arcangle=0]{d}{f}

\ncput*{\tiny 2}

\ncarc[arcangle=0]{b}{e}

\ncput*{\tiny 30}

\ncarc[arcangle=0]{c}{f}

\ncput*{\tiny 10}

\ncarc[arcangle=0]{f}{e}

\ncput*{\tiny 3}

\end{pspicture}
\label{fig:becf}}
\subfloat[Relaxation of $\left(de\right)$ and $\left(df\right)$]{
\centering

\psset{unit=50pt}

\begin{pspicture}[showgrid=false](-1,-1.25)(2.25,1.25)

\cnodeput{0}(0,0){a}{{\tiny $0,0$}}

\nput{180}{a}{$a$}

\cnodeput{0}(.7,.7){b}{{\tiny $4,8$}}

\nput{90}{b}{$b$}

\cnodeput{0}(0.7,-0.7){c}{{\tiny $1,2$}}

\nput{-90}{c}{$c$}

\cnodeput{0}(1.4,0){d}{{\tiny $10,16$}}

\nput{90}{d}{$d$}

\cnodeput{0}(2.1,0.7){e}{{\tiny $36,56$}}

\nput{90}{e}{$e$}

\cnodeput{0}(2.1,-0.7){f}{{\tiny $12,20$}}

\nput{-90}{f}{$f$}

\ncarc[arcangle=0]{a}{b}

\ncput*{\tiny 4}

\ncarc[arcangle=0]{a}{c}

\ncput*{\tiny 1}

\ncarc[arcangle=0]{d}{b}

\ncput*{\tiny 4}

\ncarc[arcangle=0]{d}{c}

\ncput*{\tiny 8}

\ncarc[arcangle=0]{d}{e}

\ncput*{\tiny 20}

\ncarc[arcangle=0]{d}{f}

\ncput*{\tiny 2}

\ncarc[arcangle=0]{b}{e}

\ncput*{\tiny 30}

\ncarc[arcangle=0]{c}{f}

\ncput*{\tiny 10}

\ncarc[arcangle=0]{f}{e}

\ncput*{\tiny 3}

\end{pspicture}
\label{fig:def}}
\subfloat[Relaxation of $ $$\left(fe\right)$.]{
\centering

\psset{unit=50pt}

\begin{pspicture}[showgrid=false](-1,-1.25)(2.25,1.25)

\cnodeput{0}(0,0){a}{{\tiny $0,0$}}

\nput{180}{a}{$a$}

\cnodeput{0}(.7,.7){b}{{\tiny $4,8$}}

\nput{90}{b}{$b$}

\cnodeput{0}(0.7,-0.7){c}{{\tiny $1,2$}}

\nput{-90}{c}{$c$}

\cnodeput{0}(1.4,0){d}{{\tiny $10,16$}}

\nput{90}{d}{$d$}

\cnodeput{0}(2.1,0.7){e}{{\tiny $18,26$}}

\nput{90}{e}{$e$}

\cnodeput{0}(2.1,-0.7){f}{{\tiny $12,20$}}

\nput{-90}{f}{$f$}

\ncarc[arcangle=0]{a}{b}

\ncput*{\tiny 4}

\ncarc[arcangle=0]{a}{c}

\ncput*{\tiny 1}

\ncarc[arcangle=0]{d}{b}

\ncput*{\tiny 4}

\ncarc[arcangle=0]{d}{c}

\ncput*{\tiny 8}

\ncarc[arcangle=0]{d}{e}

\ncput*{\tiny 20}

\ncarc[arcangle=0]{d}{f}

\ncput*{\tiny 2}

\ncarc[arcangle=0]{b}{e}

\ncput*{\tiny 30}

\ncarc[arcangle=0]{c}{f}

\ncput*{\tiny 10}

\ncarc[arcangle=0]{f}{e}

\ncput*{\tiny 3}

\end{pspicture}
\label{fig:fin}}
\caption{\textsc{Single-Pair Bellman-Ford }on a 6-vertex graph. The
costs $\left(D_{1}\left(u\right),D_{2}\left(u\right)\right)$, are
shown inside each vertex. Edges that are not drawn have weight $\infty$.}
\label{fig:BFEx }
\end{figure*}

An example of the steps of Alg. \ref{alg:SSBF} is given in Figure
\ref{fig:BFEx }. Initially, only edges between $a$, $b$, and $c$
and $a$ have finite costs, as a result of steps 4-8 of the algorithm.
Next, $n-1$ passes are executed and in each of them all edges of
the graph are relaxed. Edge $\left(bd\right)$ is relaxed first, as
seen in Figure \ref{fig:bd}. Next, the relaxation of edge $\left(cd\right)$
reduces the cost of $D_{1}\left(d\right)$ from 12 to 10. Continuing
with the algorithm, we obtain the final result in Figure \ref{fig:fin}.
Note that in this example, the result obtained after the first pass
is the final result. In general, however, the final costs may be obtained
only after all $n-1$ passes are performed.

\begin{lem}
\label{lem:BF}Given $n$, a cost function $\varphi$, and a source
$s$, after the execution of Alg. \eqref{alg:SSBF}, one has $D_{1}\left(u\right)=\varphi^{*}\left(s,u\right)$
and $D_{2}\left(u\right)=2\cost\left(p^{*}\left(s,u\right)\right)$.\end{lem}
\begin{IEEEproof}
Let $\hat{p}\left(s,u\right)=v_{1}v_{2}\cdots v_{m+1}$ be the path
that minimizes $\bar{\varphi}\left(p\right)$ among all paths $p$
between $v_{1}=s$ and $v_{m+1}=u$. Since any path $p$ has at most
$n$ vertices, we have $m\le n-1$. The algorithm makes $n-1$ passes
and in each pass relaxes all edges of the graph. Thus, there exist
a subsequence of relaxations that relax $\left(v_{1}v_{2}\right),\left(v_{2}v_{3}\right),\cdots,\left(v_{m}v_{m+1}\right)$,
in that order. The proof for the claim regarding $D_{1}\left(u\right)$
follows by invoking the path-relaxation property and the fact that
$\varphi^{*}\left(s,u\right)=\bar{\varphi}\left(\hat{p}\left(s,u\right)\right)$.
The proof for the claim regarding $D_{2}\left(u\right)$ is similar.
\end{IEEEproof}
\bibliographystyle{IEEEtran}
\bibliography{bib}

\end{document}